\documentclass[aps,twocolumn,showpacs,superscriptaddress]{revtex4}
\usepackage{graphicx,color, amsbsy}

\begin{document}

\title[Laser-BEC structure emission]{Co--propagating Bose--Einstein Condensates
  and electromagnetic radiation: Emission of
  mutually localized structures}
\author{A.~Kim}
\affiliation{Institute of Applied Physics, Russian Academy of Sciences, 603950 Nizhny Novgorod, Russia}
\author{F.~Cattani\footnote{f.cattani1@physics.ox.ac.uk}}
\affiliation{Department of Physics, Clarendon Laboratory, OX1 3PU, Oxford, UK}
\author{D.~Anderson}
\author{M.~Lisak}
\affiliation{Department of Radio and Space Science, Chalmers
University of Technology, SE-412 96 G\"{o}teborg, Sweden}

\date{\today}

\begin{abstract}

Using a semi-classical model to describe
the interaction
between coherent electromagnetic radiation and a Bose-Einstein
condensate in the  limit of zero temperature, including the back
action of the atoms on the radiation, we have analyzed the phenomenon
of emission of solitary-like wave packets which can accompany the 
 formation of mutually localized atom-laser structures.
\end{abstract}



\maketitle
\section{Introduction}
With the realization of Bose-Einstein condensates (BEC) and of
coherent atomic beams all the questions inherent to the
manipulation of such systems have acquired a certain importance. In
particular, early studies such as \cite{ref:first,ref:zhang} have
started  an  interest in the manipulation of atomic
structures via their interactions with coherent electromagnetic
radiation. 
 Not
only these studies could be of importance for applications such as
atom interferometry, they also offer a possible test of the
analogies between optics and 
quantum matter waves. In fact, BECs provide us
with a quantum system where matter waves can be realized on
macroscopic scales and which, under the approximations of zero
temperature, low densities, weak interactions and within the limits of
validity of a mean field theory, is amenable to a mathematical 
description based on the Gross-Pitaevskii equation, completely akin to
the basic equation of nonlinear optics,
i.e. the nonlinear  Schr\"{o}dinger equation, \cite{ref:review}. The
Kerr-like  nonlinearity is given for the atoms by the
atom-atom interactions.
It has been demonstrated that it is
possible to
reproduce typical optical and nonlinear optical phenomena with a 
 a BEC, from the generation of solitons
to four-wave mixing, from parametric amplification to second harmonic
generation, to mention only a few of them (for a review, see
\cite{ref:meystre} and references therein). It is possible to 
push the analogy even further and consider the electromagnetic
radiation as the medium that allows for nonlinear interactions between
atoms as discussed initially in  \cite{ref:first}. This corresponds
exactly to the optics case where the medium 
through which radiation propagates can 
bring about nonlinear effects for the electromagnetic field: 
Nonlinear effects in the dynamical
evolution of the atoms and of the radiation are then a consequence of the
atom-light interactions. Models of this
interaction were presented by several authors, \cite{ref:zhang,ref:models} and
Krutitsky {\em et al.}, \cite{ref:krutitsky}, gave a full derivation of
the equations describing it starting from first
principles within the framework of quantum field theory. This last work
showed the emergence of a
resonant nonlinear term in the system dynamics as a consequence of the laser-atom
dipole-dipole interaction, besides the well known
Kerr-like nonlinearity. 
The same
equation for the atoms and consequently a coupled system of equations for the
laser-atom system were rederived in \cite{ref:us1}, this time starting
from a semi-classical theory. It was found there that the
response of the ''medium'', that is of the laser radiation, to the
dynamics of the condensate could play an important role in the coupled
evolution and even allow for the formation of mutually localizd
atom-laser structures capable of propagating with no changes in the
atom density and laser intensity, in spite of the assumed repulsive
atom-atom interaction. Such solitary-like structures are of interest
bacause of  their
properties of self-localization and robust propagation and the effects
of these interactions 
can be seen also in relation to the creation of
meta-lenses and comoving potentials to refocus atom waves, \cite{ref:comoving}. Their emerging
as a result of the coupling during 
propagation of atoms and laser was studied in \cite{ref:us1} while the
stationary states of the coupled system and their stability properties
were introduced in \cite{ref:us2}. However, little is known about their
actual mechanism of formation. In the optics case of a focusing
nonlineariry,  we would expect an initial bell-shaped structure to
shed away the radiation which cannot 
be accomodated and to adjust asymptotically to a soliton wave. It is
interesting to see whether the same happens in the
present case of coupled atom-laser propagation and how, since the
coupling may lead to novel effects. We have therefore studied
numerically the process through which such coupled  soliton-like
objects are formed evidentiating the occurrence of a phenomenon
reminiscent of soliton emission in nonlinear optics,
\cite{ref:emission}. In fact, the equations analyzed here, predict the
formation of solitary-like structures for both atoms and light which
can move away from the region where they were generated. Although in
our case there is no external trapping but only the self-consistent
interaction of atoms and laser, these results are suggestively similar
to escaping solitons described and observed in completely different
environments, for instance in nematic liquid
crystals, \cite{ref:escaping}.\\
We will breafly review the basic physics of the
semi-classical model  and the limitations to be
considered in Sec.\ref{sec:basics}. Sec.\ref{sec:initial} presents an
investigation of the initial evolution of the coupled system which
will than be studied numerically in Sec.\ref{sec:numerics}.

\section{Semi-classical model and set up of the problem}\label{sec:basics}

The basic physics of atom-laser interactions in the simplest dipole 
approximation is given by photons exciting atoms which in turn re-emit
photons absorbed by other atoms, thus giving rise to a long-range
interatomic interaction, \cite{ref:cohen}. Details of the
semi-classical derivation of both the atom and the laser equations are
given in \cite{ref:us1, ref:us2}, here we will only briefly review
the two model equations and re-introduce the notation. In a semi-classical
derivation, the force exerted by the light on the atoms is written as
the gradient of a potential and this potential is used as the
atom-laser interaction term in the
Hamiltonian for the atoms Schr\"{o}dinger equation.   Such
force term is a generalization of the ponderomotive force and takes
into account the possibility of an inhomogenoeus gas. The existence
of stationary solutions is physically crucial so we will study
a far-off resonant monochromatic field
 ${\bf{E}}({\bf{r}}, t) = Re[{\boldsymbol{\mathcal{E}}} ({\bf r})\exp(-i\omega_L t)]$ (where
 ${{\boldsymbol{\mathcal{E}}} ({\bf r})}$ is the  
complex amplitude of the laser field). The time  averaged
force (over
laser cycles)  is  
$ {\bf F}=\frac{1}{16\pi}\nabla\left[|{\bf{E}}|^2
  \frac{\partial\epsilon}{\partial n}\right] = -{\boldsymbol{\nabla}}
V_d. $
Here $\epsilon(\omega, n)$ is the medium dielectric constant with
atom density $n$ and is given by $  \epsilon(\omega, n)=1+\frac{4\pi\alpha n}{1-\frac{4\pi}{3}\alpha n},$
where, as derived from quantum theory,
$\alpha(\omega)=-d^2/\hbar\Delta$ is the atomic  polarizability at the
laser frequency $\omega_L$, with $\Delta=\omega_L -\omega_a$ being the
detuning from the nearest atomic resonance frequency $\omega_a$, and
$d$ is the dipole matrix element of the resonant
transition,\cite{ref:krutitsky,ref:born}. The relative semplicity of
the semi-classical derivation comes at the price of restricting the
validity of the model to a well defined range of parameters: The
concept of force is purely 
classical, therefore quantum fluctuations, stochastic heating and any
incoherent process are to be neglected. This limits the validity of
this model to  large detunings $|\Delta| \gg
\omega_a, \Gamma$ ($\Gamma$ is the atoms natural line
width). Under these limitations, the potential $V_d$
can be inserted into the
atom Gross--Pitaevskii equation where it describes the laser-induced dipole-dipole
interaction between the atoms:
\begin{equation}\label{eq:Psi}
   i\hbar\frac{\partial\Psi}{\partial t}=
   \hat{H}_0\Psi
   +\left[U_0|\Psi|^2-
   \frac{\alpha}{4}\frac{|{\bf{E}}|^2}{\left(1-\frac{4\pi}{3}\alpha
   |\Psi|^2\right)^2}\right]\Psi.
\end{equation}
Here $\hat{H}_0$ is the linear single-particle Schr\"{o}dinger
Hamiltonian, the wave function $\Psi$ is normalized as
$N=\int|\Psi|^2d{\bf r}$ with $N$ denoting the total number of atoms,
so that the gas density is $n=|\Psi|^2$, $U_0=4\pi\hbar^2a_s/m$, $m$
is the atom mass and $a_s$ is the $s$--wave scattering length
(which will be assumed positive as for repulsive atom-atom interactions).
Furthermore, since
we are interested in the stationary behaviour of the system and we
have already assumed a stationary form for the electromagnetic field,
we will consider $\Psi({\bf{r}}, t) = \Phi({\bf{r}})\exp(-i\omega_a t)$.
 The atom
equation was already derived in \cite{ref:krutitsky} within a fully 
quantum model and it is
important to underline that, once the limitations of the
semi-classical reasoning are taken into account, the two derivations
lead to the same equation.\\
To describe the role played by the electromagnetic field and the
effect of the atoms on such field,  it is
necessary to include a field equation. Maxwell's equations for
the propagation of radiation in a medium, \cite{ref:born,
  ref:jackson, ref:us1}, yield a wave equation
which, under the assumption of $L_n \gg \lambda_L$ or
${\boldsymbol{\nabla}} \epsilon \cdot {\bf{E}} \simeq 0$ ($L_n$ is the
characteristic length scale of transverse density modulations and
$\lambda_L$ is the radiation wavelength), gives the three scalar
equations  ($\omega_L = k_L c$)
\begin{equation}\label{eq:A}
\nabla^2 {\boldsymbol{\mathcal{E}}} + k_L^2 \left(1 + \frac{4\pi\alpha
    |\Phi|^2}{1-\frac{4\pi}{3}\alpha|\Phi|^2}\right){\boldsymbol{\mathcal{E}}}=0.
\end{equation}
When the input field distributions do no match the
exact stationary solutions (which can be found numerically,
\cite{ref:us2}),  propagation effects of some sort are to be
expected. As demonstrated in \cite{ref:us1}, in the case of {\em red
detuning}, the system settles down asymptotically  to a
stationary state with mutually localized atom-laser structures:
Starting from a gaussian atom density profile and a 
super-gaussian laser intensity one, the interaction leads to the
formation of two bell-shaped structures which propagate unchanged
thereon. This means that
atoms and radiation in excess will be shed away,
which is the process we would like to elucidate here. Choosing $z$ as the propagation coordinate and 
limiting the investigation to slow envelope variation, we consider
\begin{eqnarray}
{\boldsymbol{\mathcal{E}}}({\bf{r}}) &=& a(x, z) \exp (ik_L z) {\bf{e}},\\
\Phi({\bf{r}}) &=& \psi(x, z) \exp (ik_a z),
\end{eqnarray}
where $x$ denotes the dimension transverse to
 the propagation direction $z$ (one transverse dimension
 only for simplicity), ${\bf e}$ is the polarization vector
 of the field  and $k_a$ is the atom
wave number.  The coupled system of equations (\ref{eq:Psi}),
(\ref{eq:A}), can then be written in normalized variables as
\begin{equation}\label{eq:atom}
i \mu\frac{\partial \tilde{\psi}}{\partial  \tilde{z}} = -\frac{1}{2} \frac{d^2
 \tilde{\psi}}{d \tilde{x}^2} + \frac{1}{2}\beta_{coll}|\tilde{\psi}|^2\tilde{\psi}
-\frac{s}{2}\frac{|\tilde{a}|^2}{\left(1-s|\tilde{\psi}|^2\right)^2}\tilde{\psi}
\end{equation}
\begin{equation}\label{eq:laser}
i\frac{\partial \tilde{a}}{\partial \tilde{z}} = -\frac{1}{2}\frac{d^2
  \tilde{a}}{d \tilde{x}^2}-\frac{3s}{2}\frac{ |\tilde{\psi}|^2  \tilde{a}}{1-s|\tilde{\psi}|^2},
\end{equation}
where the following normalisation has been used: $\tilde{x} = {x}k_L$, for the
atom wave function $ \tilde{\psi} = \psi/\psi_*$ with
$  (4\pi|\alpha|/3)
\psi_*^2 = 1$, for the laser $\tilde{a} =
a/a_*,$ with $ m|\alpha| a_*^2/(2\hbar^2
k_L^2) = 1$, $s = sign(\alpha)$, $\mu = k_a/k_L$ (for simplicity we
will assume $\mu = 1$ hereafter) and $\beta_{coll} =
6a_s/(k_L^2 |\alpha|)$. The tilde will be dropped hereafter unless
otherwise stated. The red detuning case studied here will correspond
to $s =+1$. Notice that no mutual localization is possible in the
blue detuning case. 
While the classical description for the laser field is
justified by the choice of the intensity regime, 
for a mean field model to be valid for the atom wave function, we must
consider not only a zero temperature limit but also 
a low density limit with $n a_s^3 \ll 1$, see
\cite{ref:review}. Furthermore, a low density regime is required in
order to avoid the singularity of the model and consequent spurious
collapse-like phenomena.

\section{Initial evolution}\label{sec:initial}

As done previously, we will start assuming an initial laser
intensity profile in super-gaussian form much wider than  the gaussian
initial atom 
density profile, both of them definitely different from the stationary
solutions of the system thus ensuring a dynamical evolution:
\begin{eqnarray}
\psi(x,0) &=& \psi_0 e^{\left(-x^2/2 d_a^2\right)}\label{eq:initialpsi}\\ 
a(x,0) &=& a_0 e^{-\left(x^2/2 d_L^2\right)^g}\label{eq:initiala}
\end{eqnarray}
where $g$ is the supergaussian parameter ($g =
10$ in the simulations). The flat-top laser profile
eliminates gradient forces on the atoms at the very initial
stage. However, the flat top is immediately modified due to
the natural evolution of the system and the initial steps will be the
seed of the subsequent structure generation. 
 The atoms will imprint a chirp on the laser with
the effect of creating a central intensity peak with two lateral
throughs, \cite{ref:us1}. This can be formally seen via a perturbative solution of
the first propagation stage (i.e. for $z\ll \lambda_L$). With
$|\psi|^2 \ll 1$, 
the denominators in Eqs.(\ref{eq:atom}) and (\ref{eq:laser})
can be expanded keeping terms up to 
the order $\sim|\psi|^2$.  Separating amplitude and phase as
$a(x,z) = A(x,z)\exp(i\theta(x,z)), \psi(x,z) = B(x,z)\exp(i\phi(x,z))$
 the two equations give, upon separation of real and imaginary parts,
\begin{equation}
\left\{
\begin{array}{l}
 \frac{\partial \phi}{\partial z} =
\frac{1}{2}\left[\frac{1}{B}\frac{\partial^2 B}{\partial x^2} - \left(
    \frac{\partial\phi}{\partial x}\right)^2\right] + \frac{A^2}{2} -
B^2\left(\frac{\beta}{2} - A^2\right),\\
\frac{\partial B^2}{\partial z} = - \frac{\partial}{\partial x}\left(B^2
  \frac{\partial\phi}{\partial xi}\right),\\
 \frac{\partial\theta}{\partial z} =
\frac{1}{2}\left[\frac{1}{A}\frac{\partial^2 A}{\partial x^2} - \left(
    \frac{\partial\theta}{\partial x}\right)^2\right] + \frac{3}{2}B^2,\\
\frac{\partial A^2}{\partial z} = - \frac{\partial}{\partial x}\left(A^2
  \frac{\partial\theta}{\partial x}\right).
\end{array} \right.
\end{equation}

Consider a perturbative expansion $F(x,z) = F_0(x) + F_1(x)
z+F_2(x) z^2$ and $G(x,z) = G_1(x)
z+G_2(x) z^2$ up to second order in $z$ where $F$ stands for the
functions $A$ and $B$ while $G$ stands for $\theta$ and $\phi$ and the
zero-th order terms are the initial functions (\ref{eq:initialpsi})
and (\ref{eq:initiala}). 
Identifying powers of $z$,  a solution is obtained for the amplitudes:
\begin{eqnarray}
&& A^2 = A_0^2(x) \left[ 1-\frac{3}{2}
  \frac{B_0^2(x)}{d_a^2}\left(\frac{2 x^2}{d_a^2}-1\right)z^2\right],\\
&& B^2 = B_0^2(x) \left[ 1+\frac{1}{2}
  \frac{\beta^\prime(x)}{d_a^2}B_0^2(x)\left(\frac{2
  x^2}{d_a^2}-1\right)z^2 - \frac{z^2}{2d^4}\right]\nonumber\\
\end{eqnarray}
where $\beta^\prime(x)=\beta-2A_0^2(x)$. This solution has the
features observed in the initial evolution of the coupled 
system: The laser intensity profile changes in such a way as to peak in the
center and at the same time two troughs are created on each side of the rising
peak. The atom density profile shows the well known nonlinear defocusing
behaviour - the center is depressed and two humps are created on both sides
of the depression. This is the beginning of the creation of the stable
mutually localized structures  discussed in \cite{ref:us1}, in a soliton-like
process the nonlinearity in the atom equation can act as a self-generated trapping
potential for the BEC.

\section{Structure emission}\label{sec:numerics}

\begin{figure}
\centering
\includegraphics[scale=0.22]{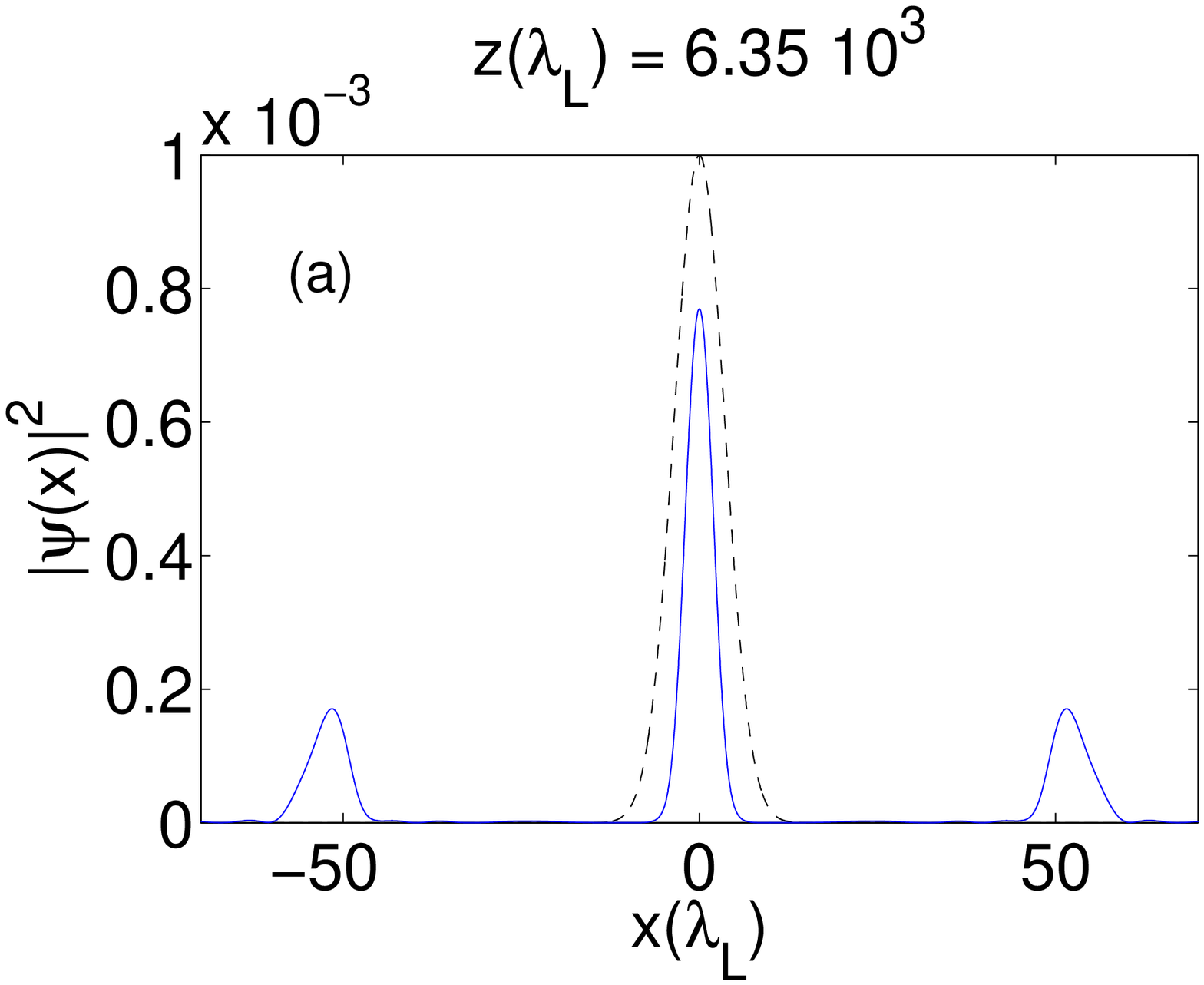}\includegraphics[scale=0.22]{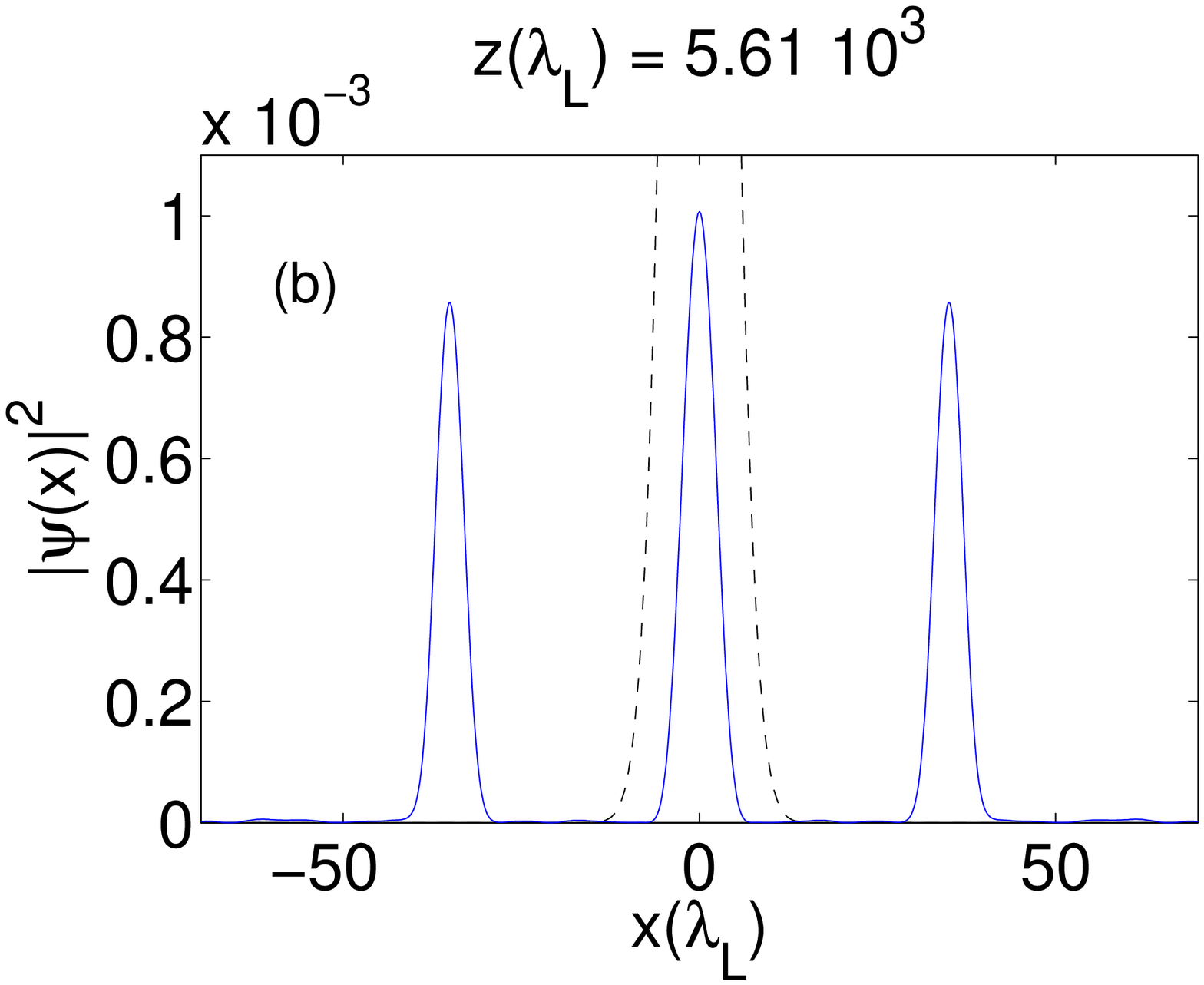}
\caption{Structure formation for (a) $\psi_0 = 0.0316$ (corresponding to
  an initial peak atom density $n_0 = 1.7 \, 10^{19}\textrm{ m}^{-3}$)
  and (b) $\psi_0 = 0.06645$ (corresponding to $n_0 = 7.51 \,
  10^{19}\textrm{m}^{-3}$). Dotted line: initial density
  distriubution. The propagation 
  distance is indicated on the plots. All other parameters as
  specified in the text. All quantities normalized as in the text.
}  \label{fig:run30020019_3102130301}
\end{figure}

\begin{figure}
\centering
\includegraphics[scale=0.22]{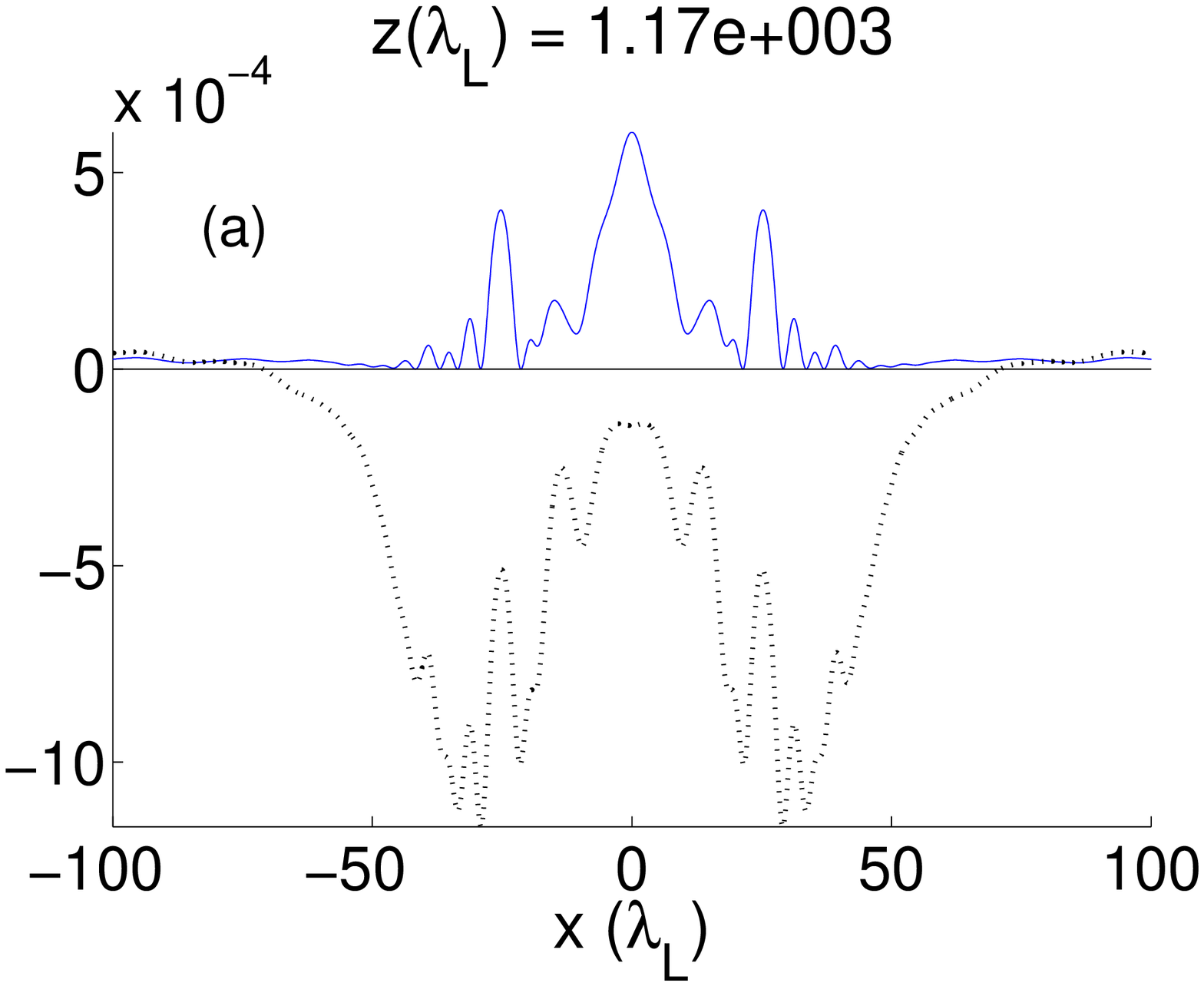}\includegraphics[scale=0.22]{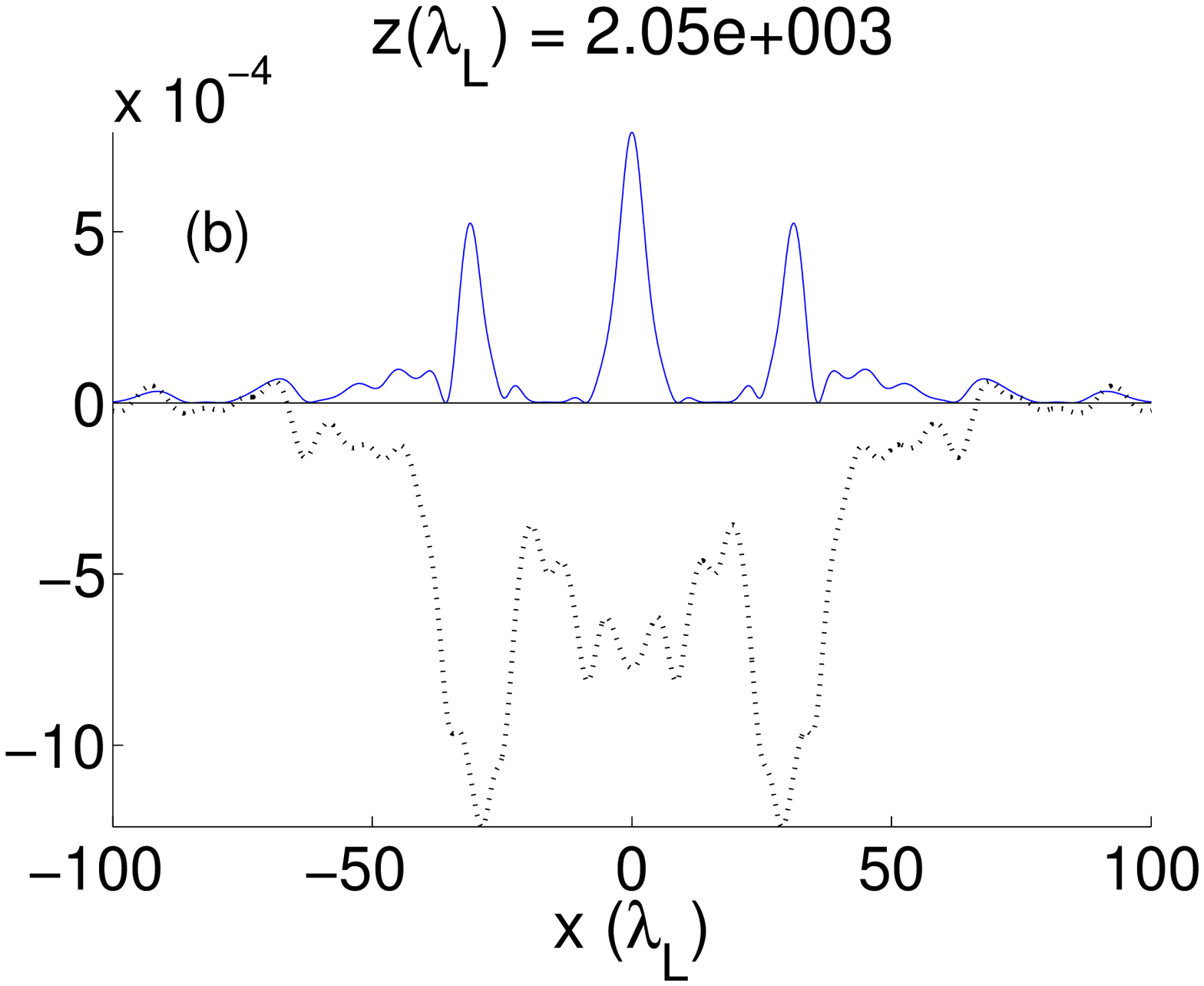}\\
\includegraphics[scale=0.22]{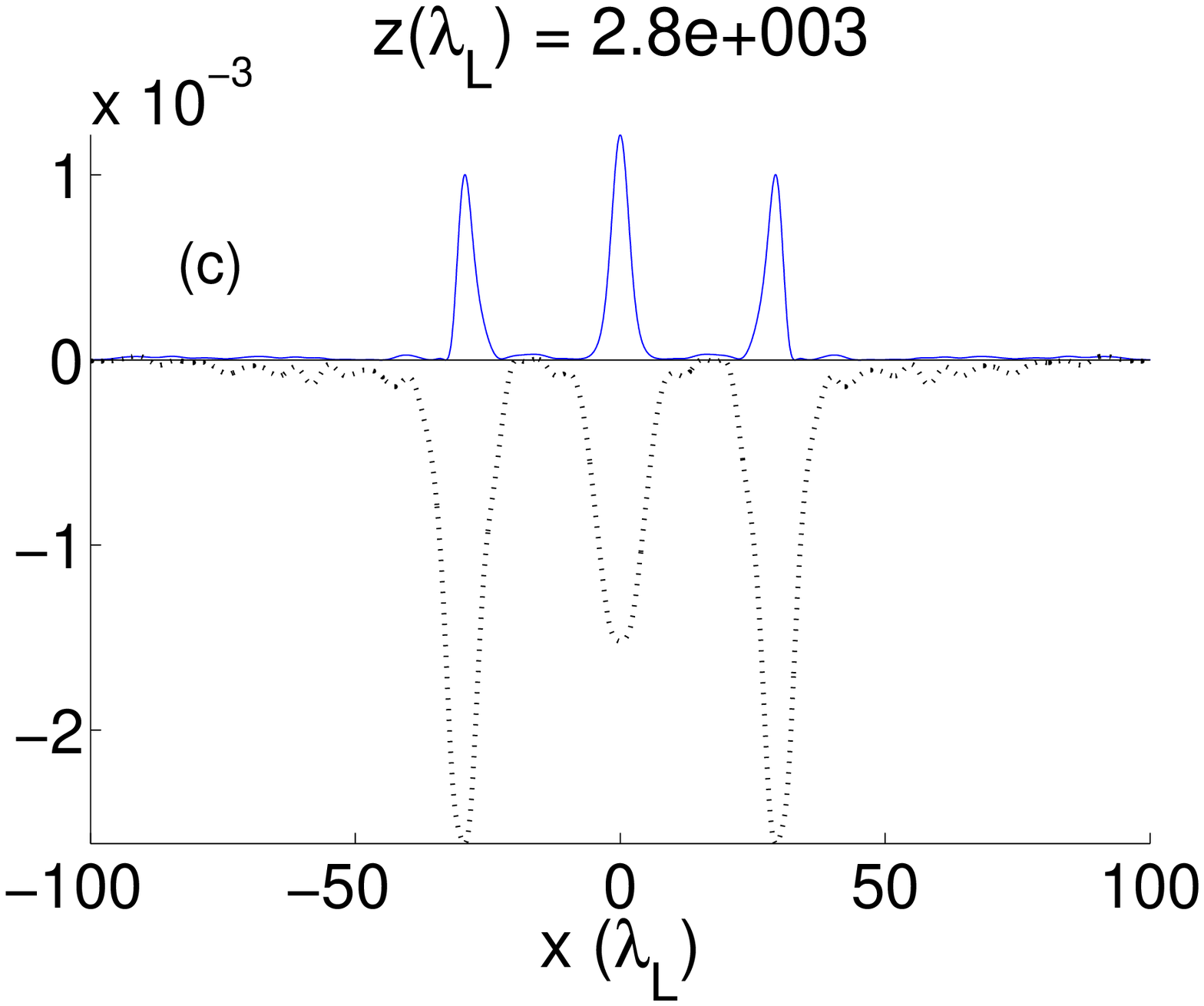}\includegraphics[scale=0.22]{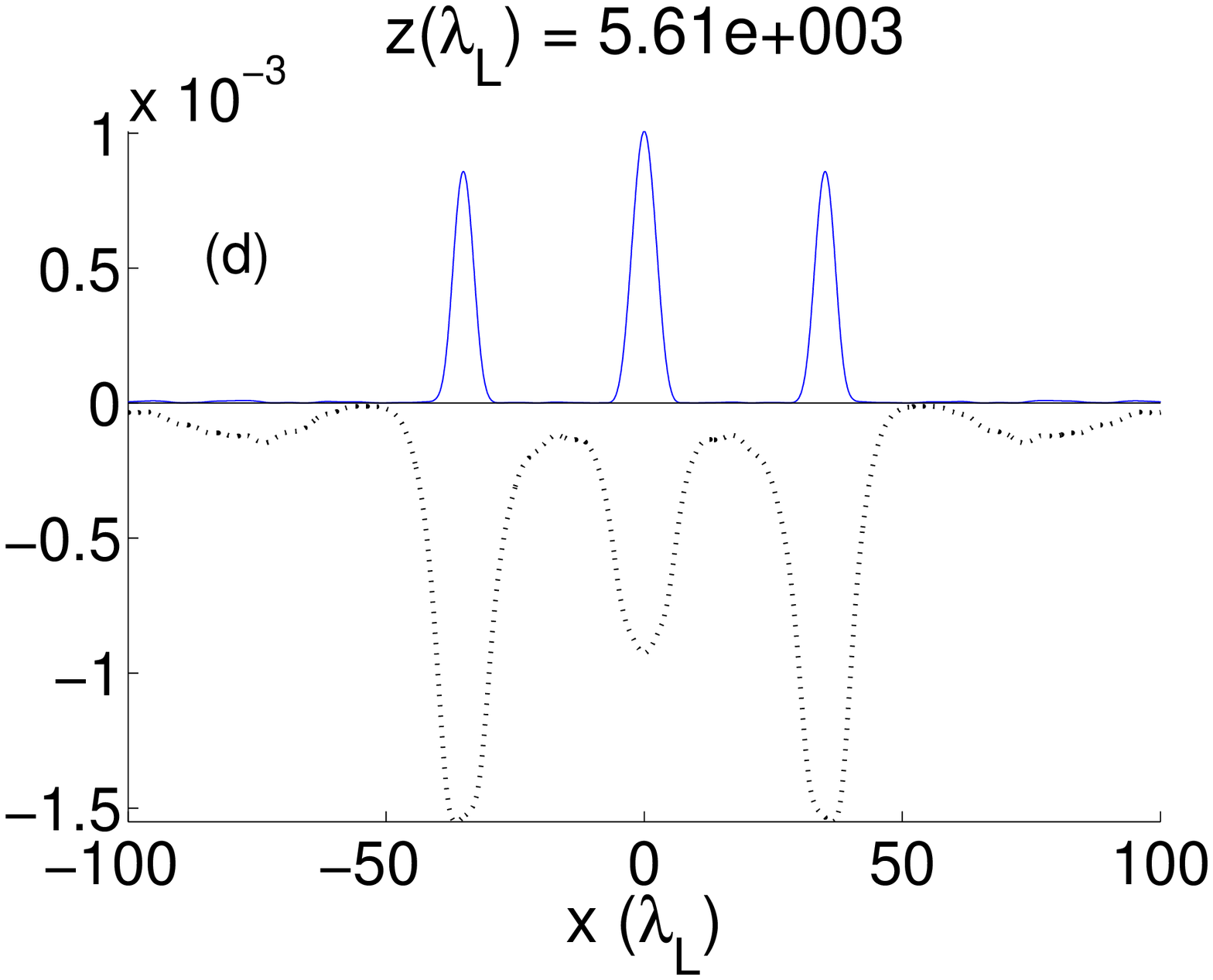}\\
\includegraphics[scale=0.22]{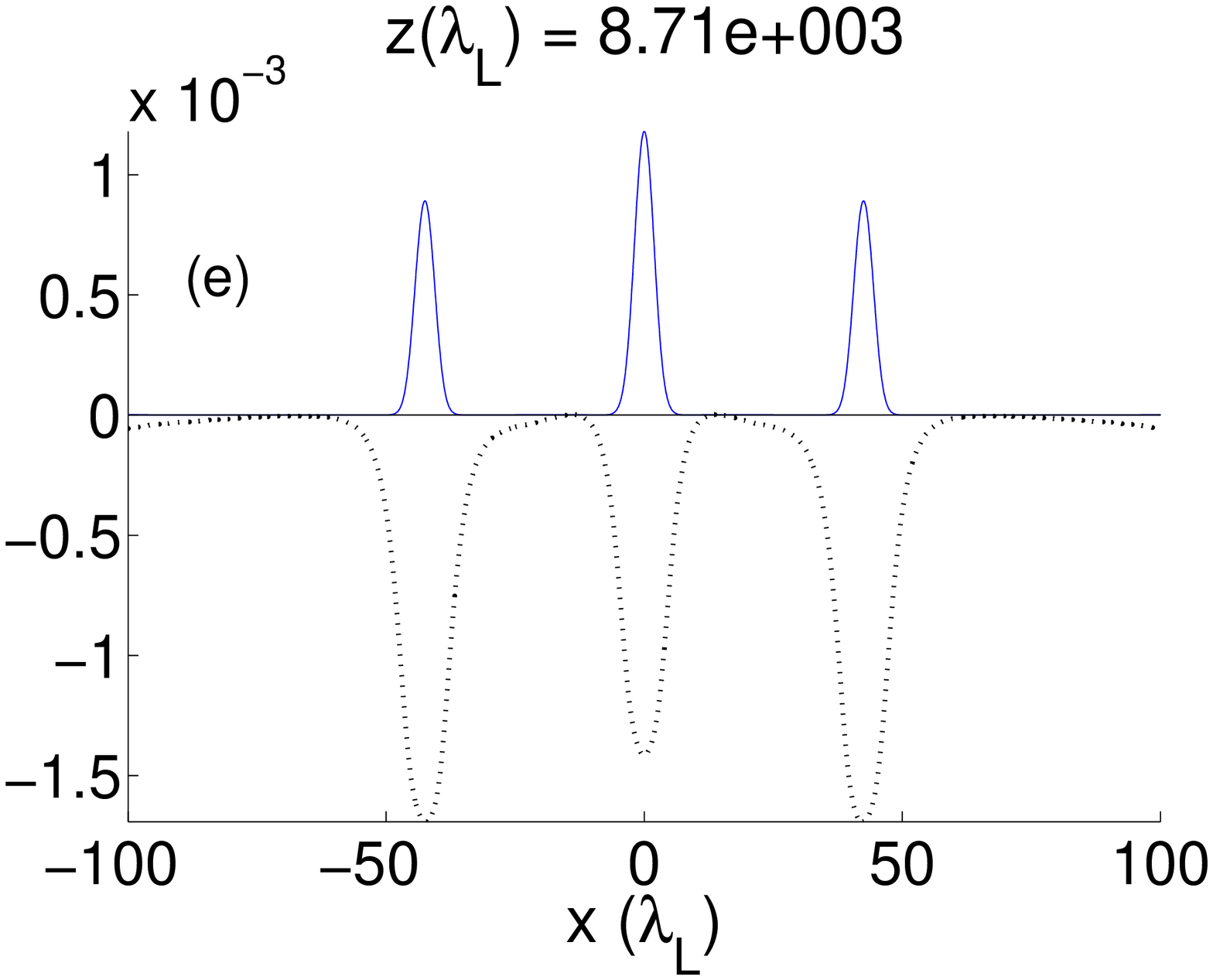}\includegraphics[scale=0.22]{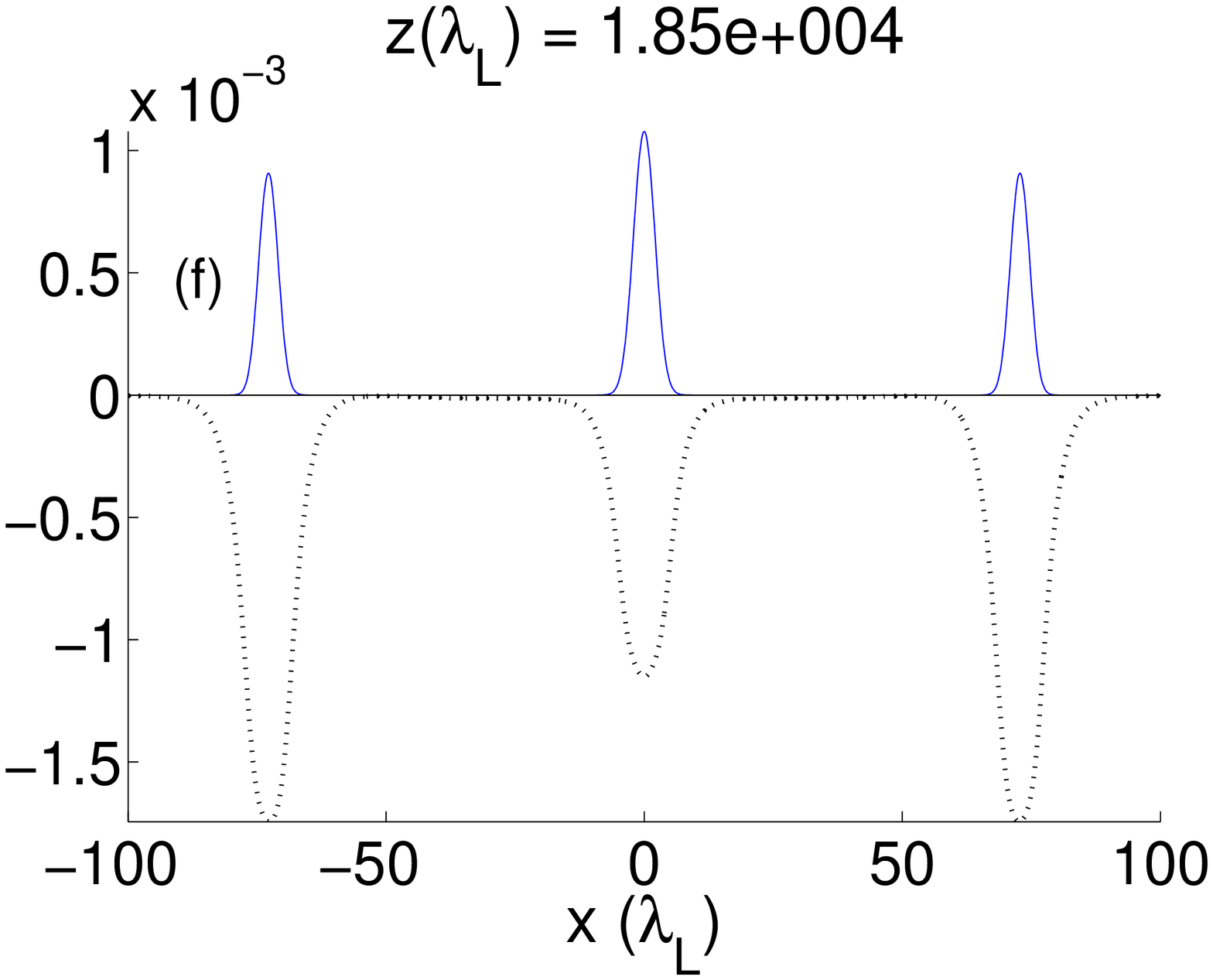}
\caption{Details of the process of structure formation for $\psi_0 = 0.06645$ ($n_0 = 7.51 \,
  10^{19}\textrm{m}^{-3}$).  Solid line: atom wave function, dotted line:
  laser-induced potential acting on the
  atoms (divided by 10 to make the figure more easily readable). The propagation 
  distance is indicated on the plots. All other parameters as
  specified in the text. All quantities normalized as in the text.
}  \label{fig:run3102130301_details}
\end{figure}

\begin{figure}
\centering
\includegraphics[scale=0.22]{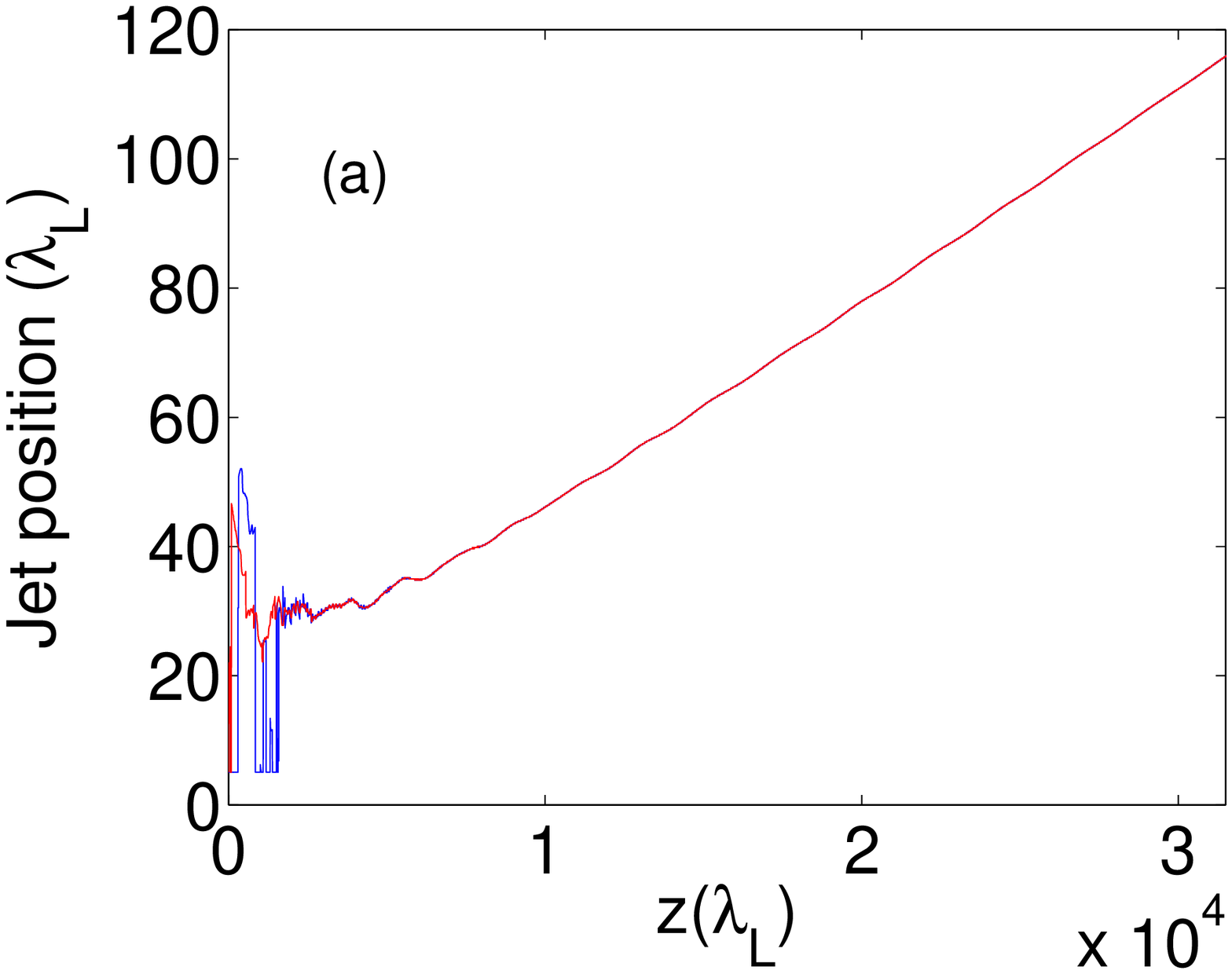}\includegraphics[scale=0.22]{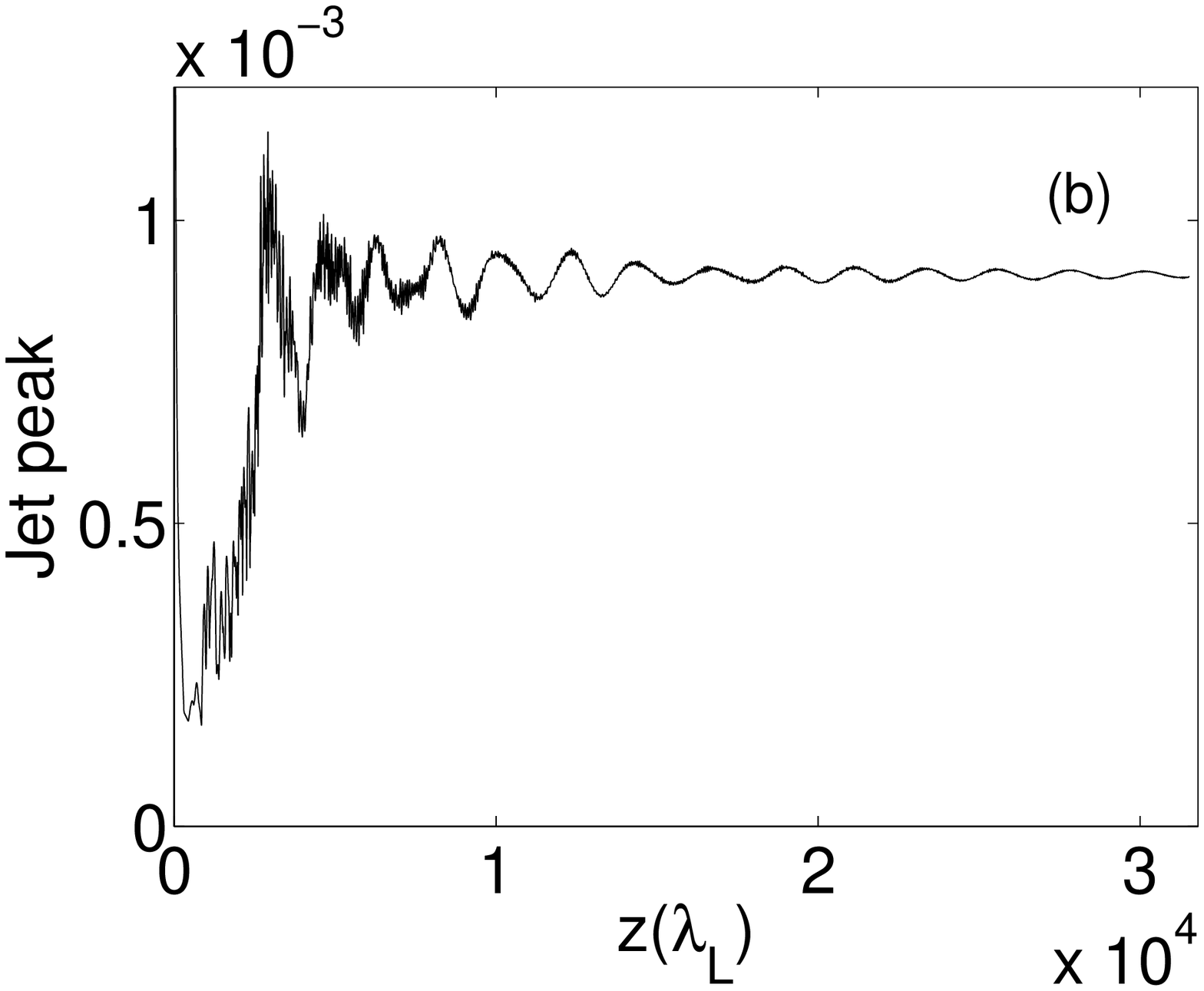}\\
\caption{(a) Emitted structure position versus propagation distance for $\psi_0 = 0.06645$ ($n_0 = 7.51 \,
  10^{19}\textrm{m}^{-3}$).  Red: laser jet position; blue: atom jet
  position. (b) Peak atom density of the emitted
  structures for the same case. All quantities normalized as in the
  text. (Color
  on line.)
}  \label{fig:run3102130301_jetposition}
\end{figure}

\begin{figure}
\centering
\includegraphics[scale=0.4]{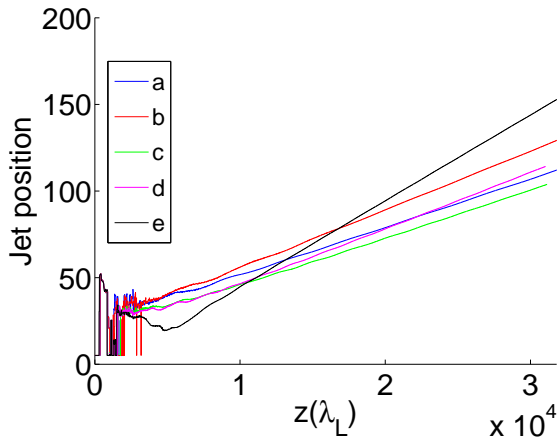}
\caption{Jet positions for different initial values of the atom peak
  density $\psi_0$. All other parameters are the same as for
  Fig.\ref{fig:run30020019_3102130301}. (a) $\psi_0 = 0.052$, (b)
  $\psi_0 = 0.054$, (c) $\psi_0 = 0.0662$, (d) $\psi_0 = 0.0664$, (e)
  $\psi_0 = 0.0668$. All quantities normalized as in the text. (Color
  on line.)
}  \label{fig:velocities}
\end{figure}

\begin{figure}
\centering
\includegraphics[scale=0.22]{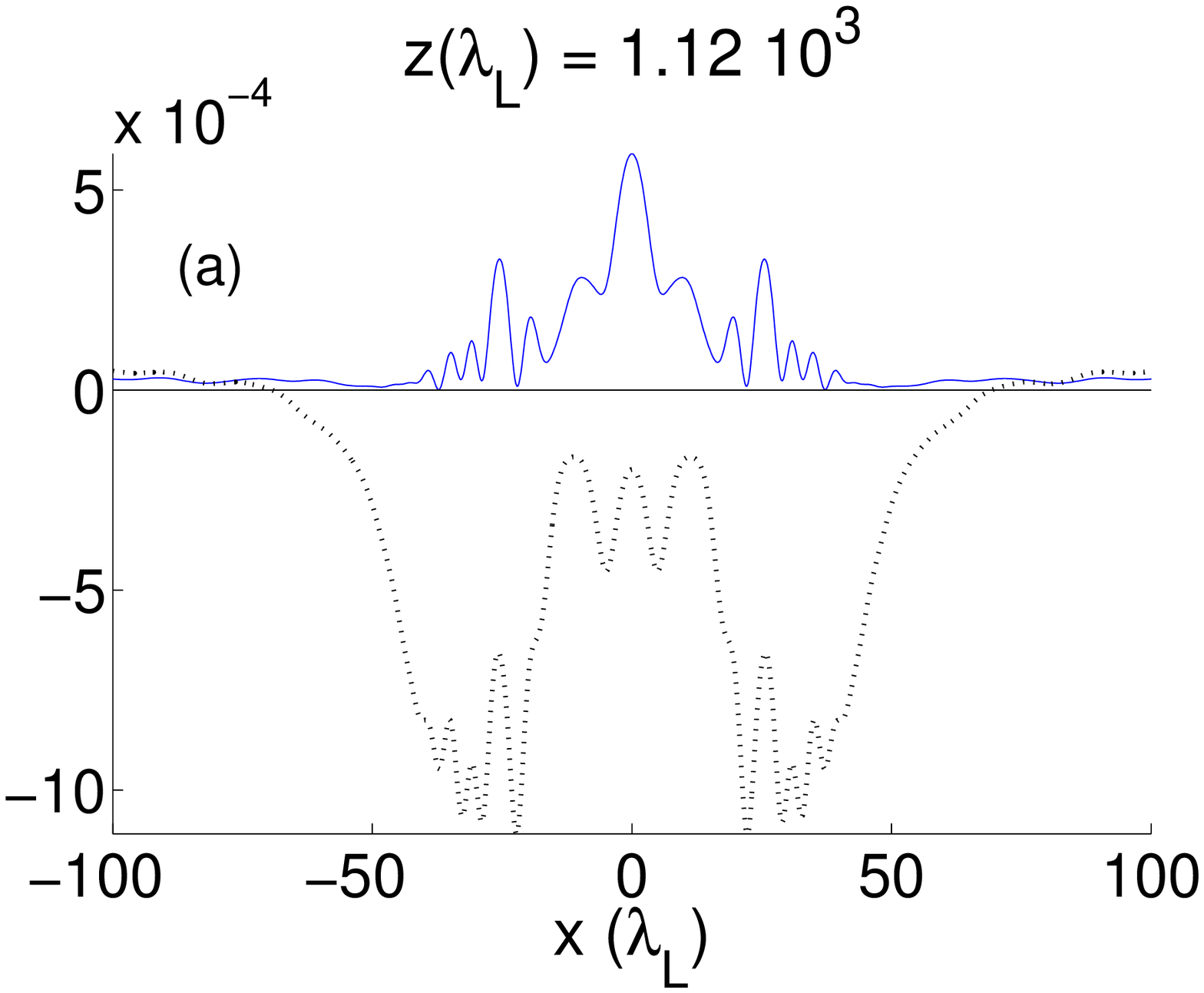}\includegraphics[scale=0.22]{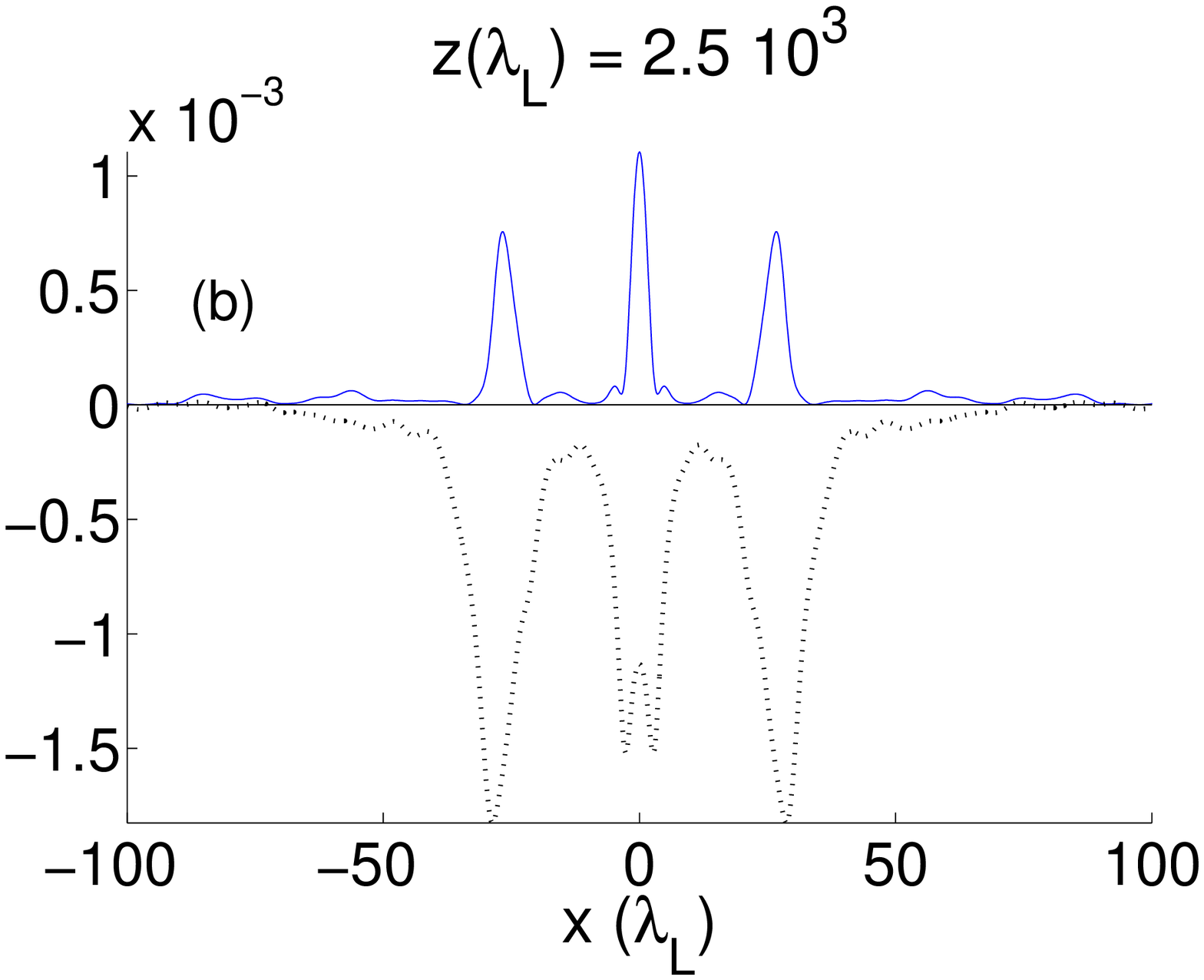}\\
\includegraphics[scale=0.22]{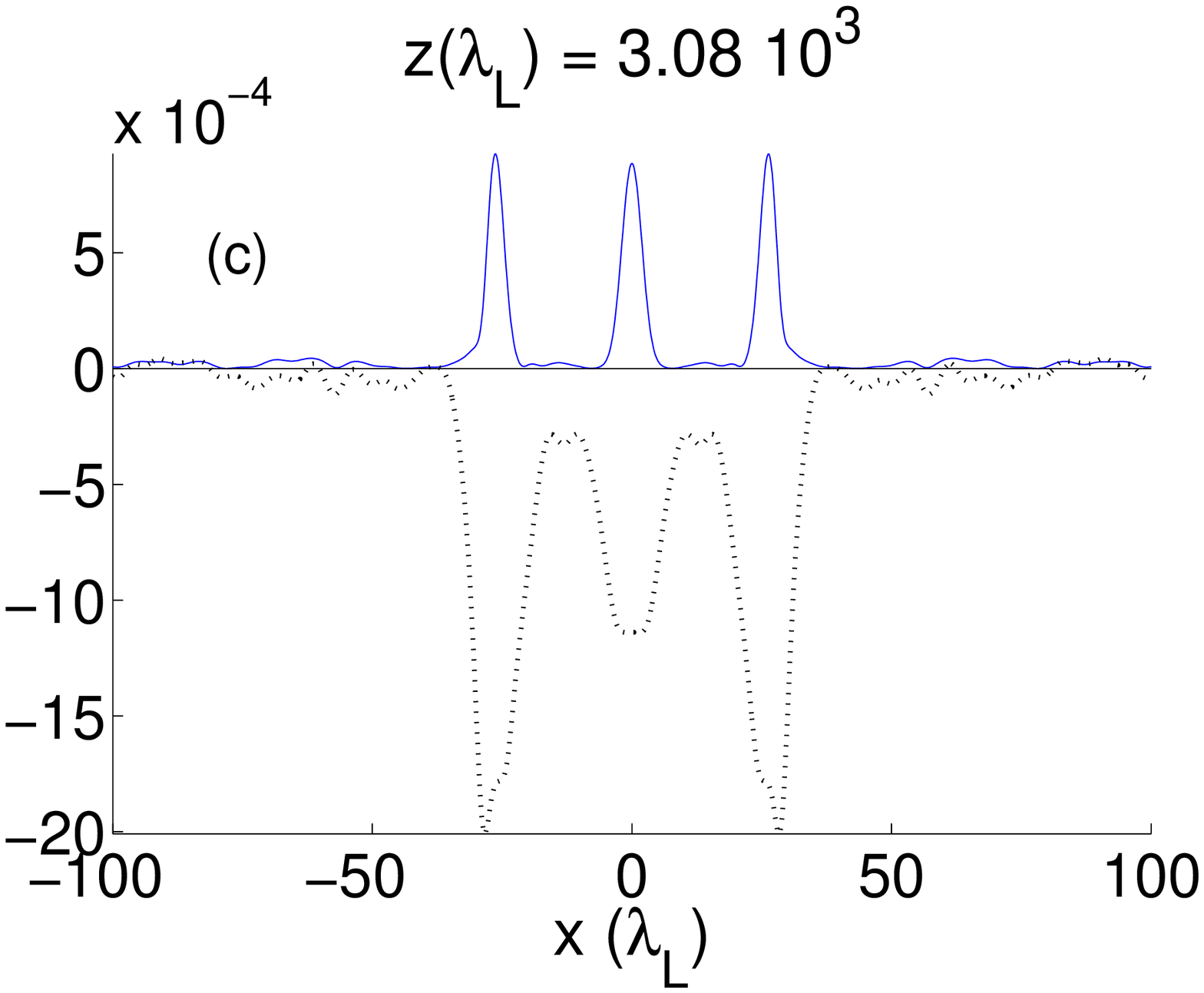}\includegraphics[scale=0.22]{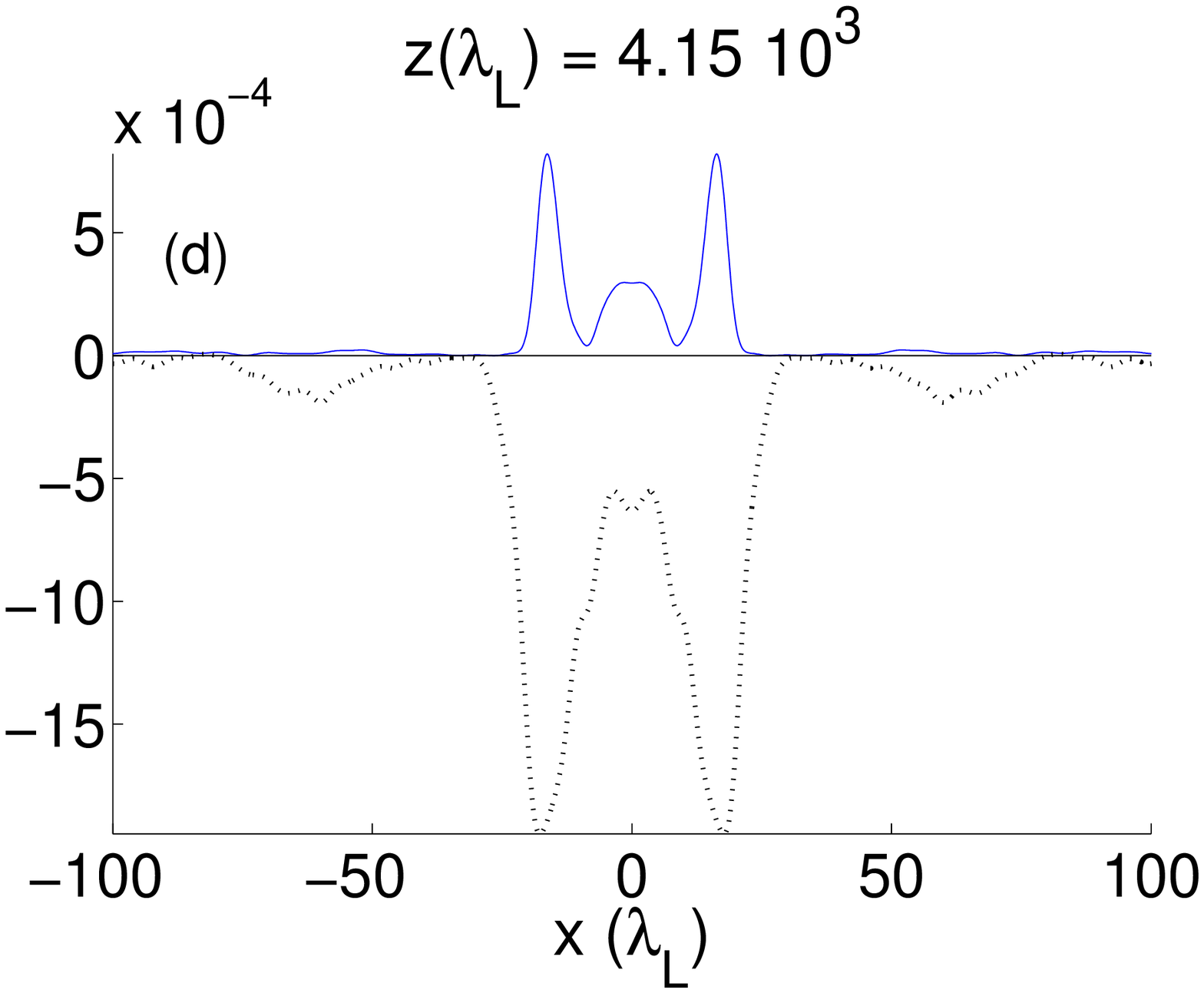}\\
\includegraphics[scale=0.22]{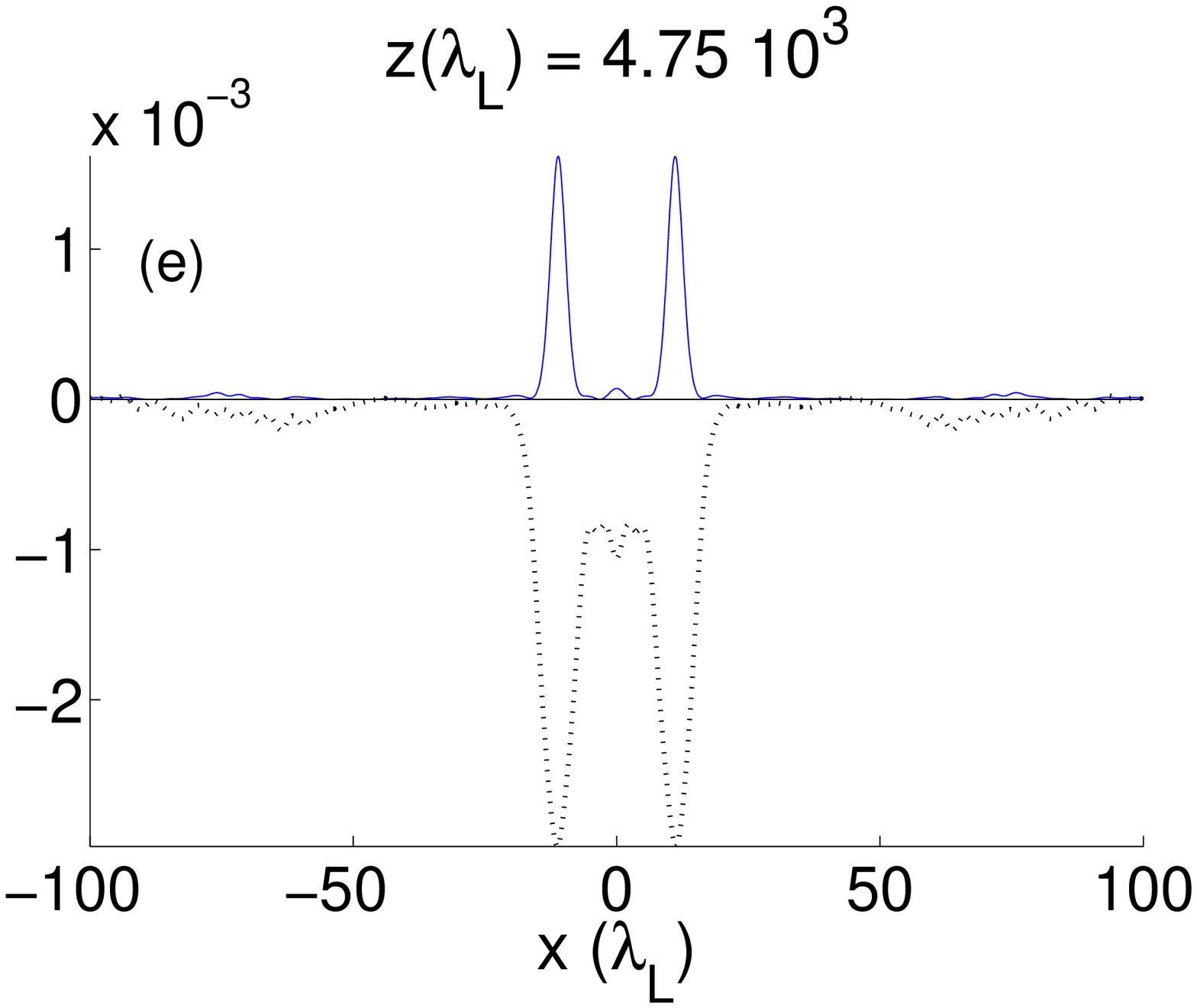}\includegraphics[scale=0.22]{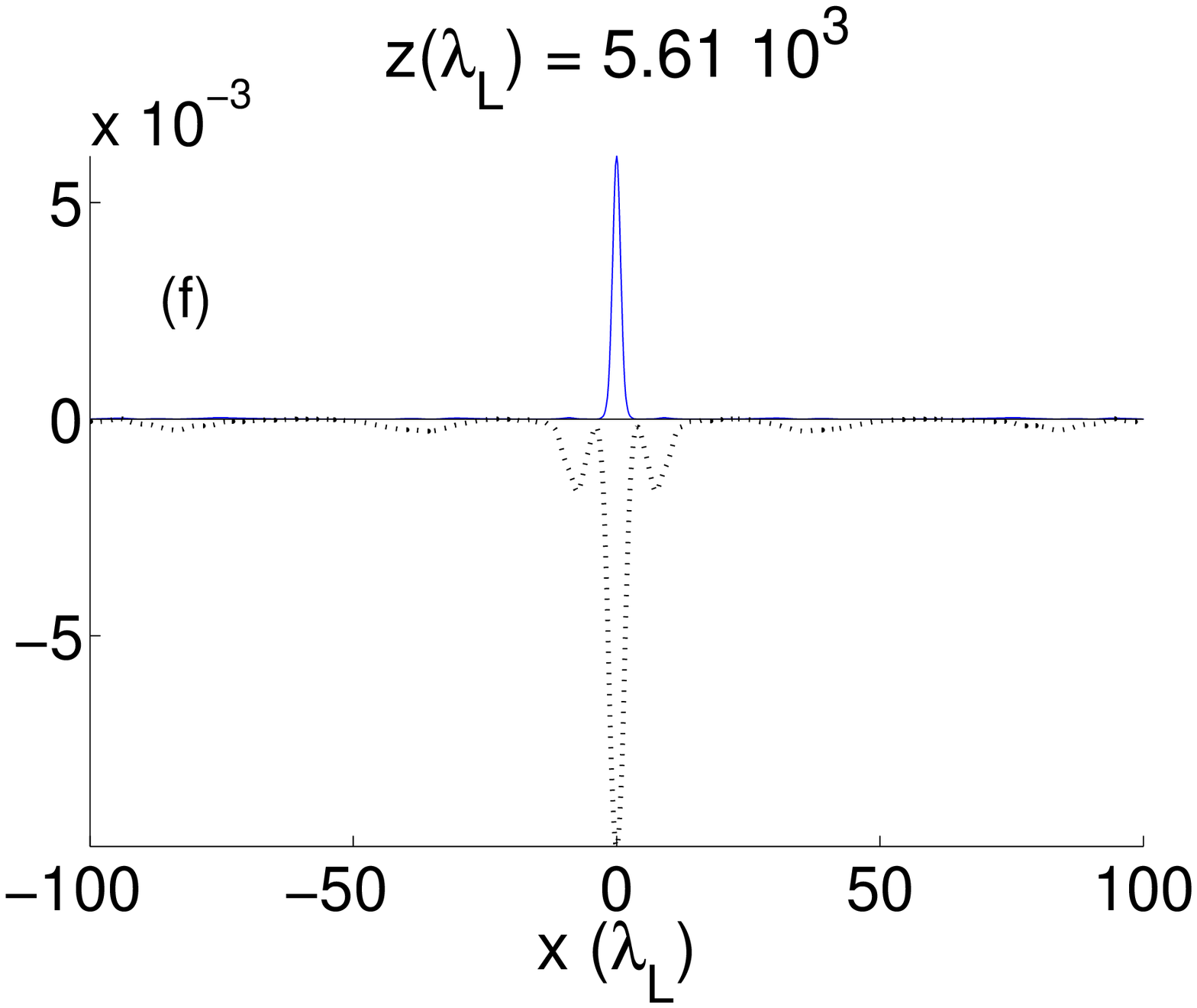}
\caption{Propagation and fusion for 
  $\psi_0 = 0.0669$ ($n_0 = 7.62 \,
  10^{19}\textrm{m}^{-3}$).  Solid line: atom wave function, dotted line:
  laser-induced potential acting on the
  atoms (divided by 10 to make the figure more easily readable). Propagation distance as indicated on the
  plots. All quantities normalized as in the text. 
}  \label{fig:run3102130601}
\end{figure}

\begin{figure}
\centering
\includegraphics[scale=0.22]{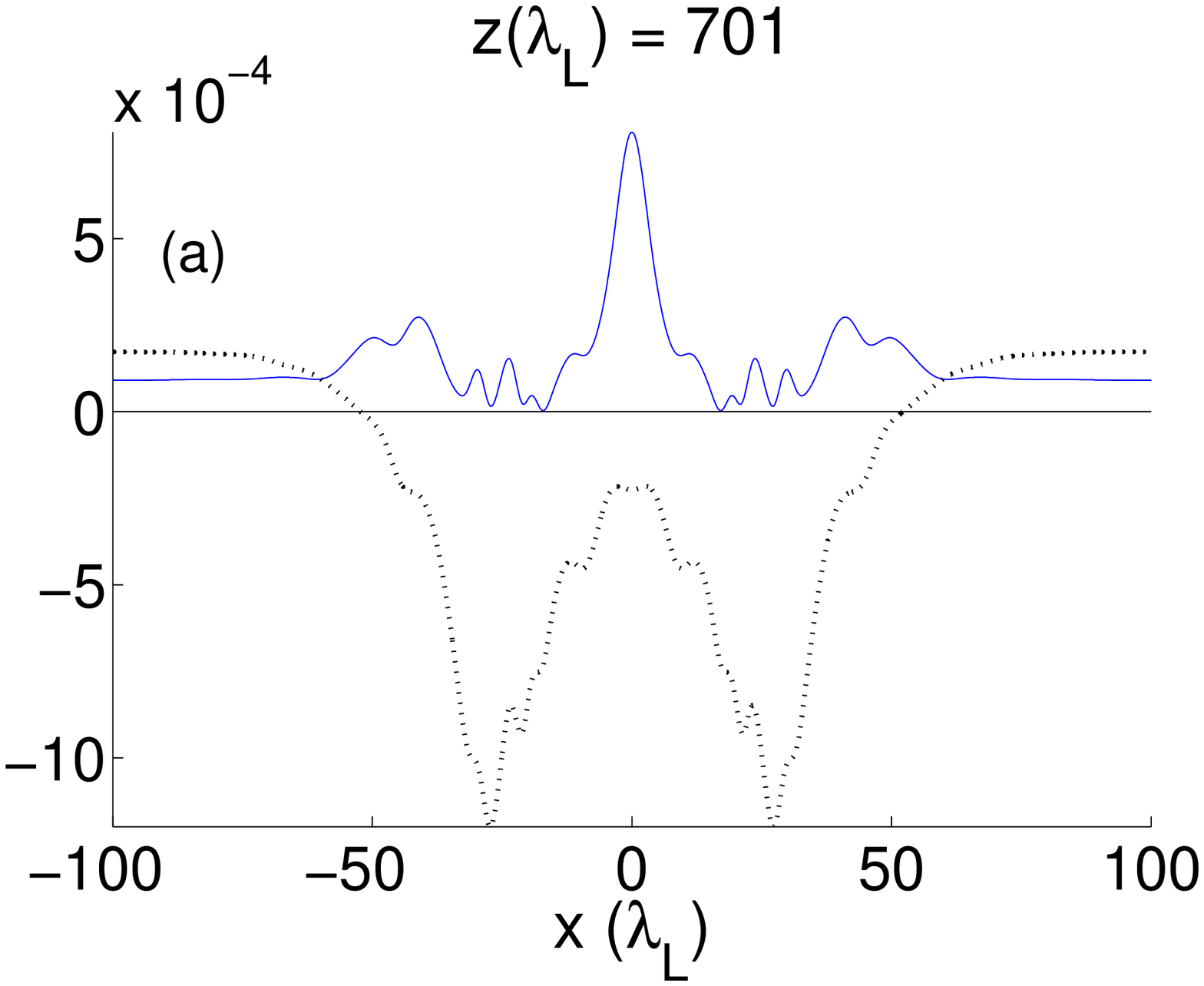}\includegraphics[scale=0.22]{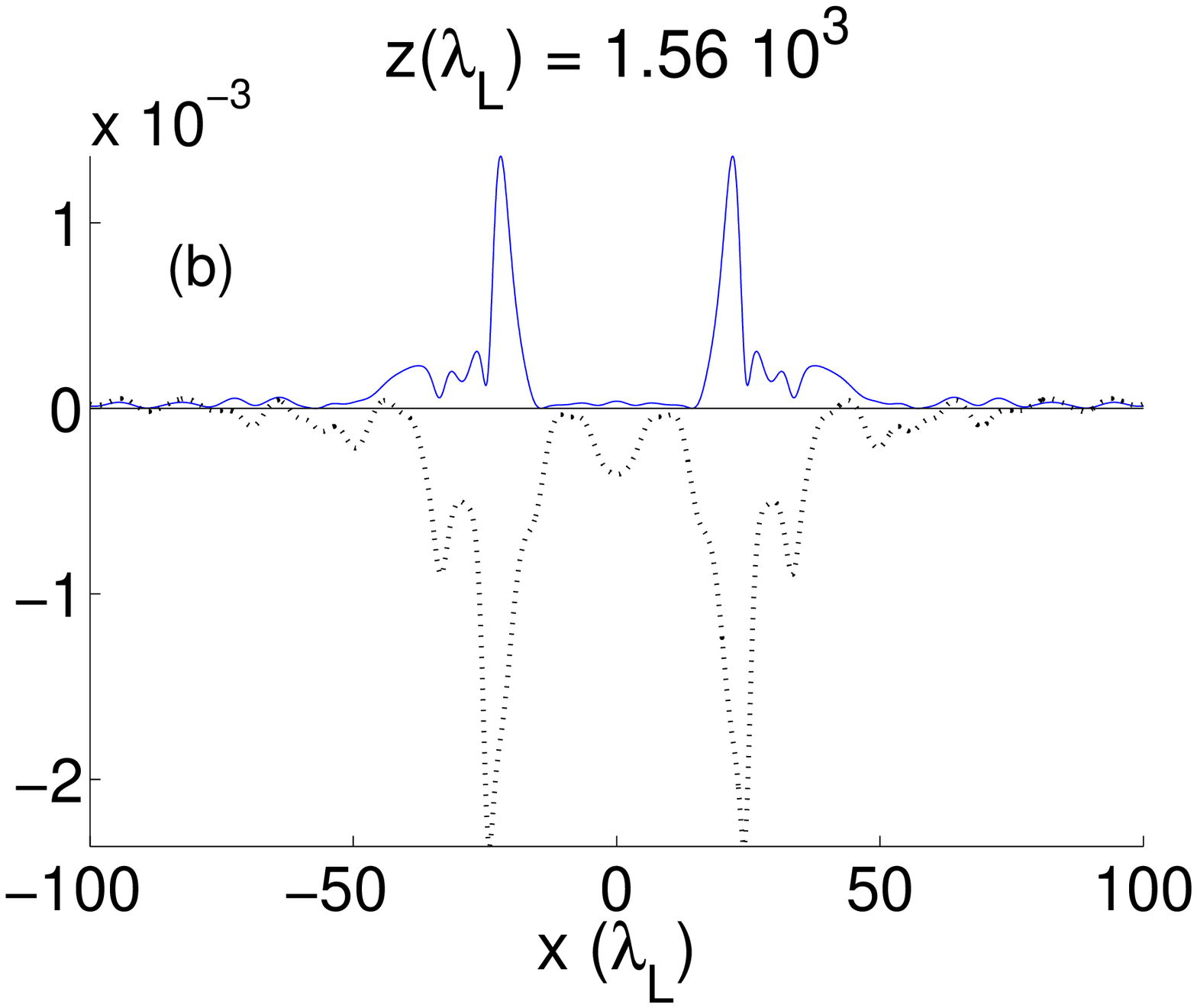}\\
\includegraphics[scale=0.22]{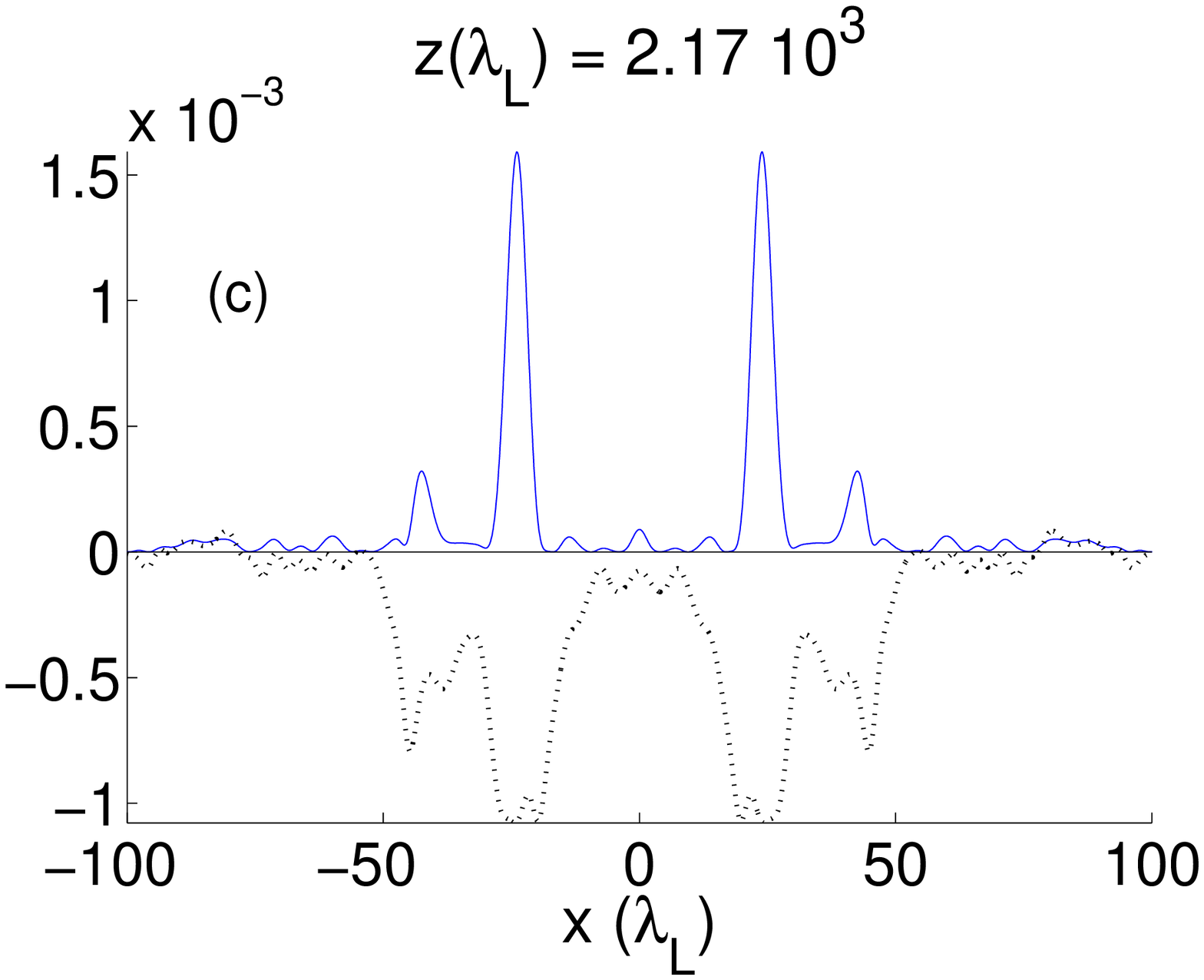}\includegraphics[scale=0.22]{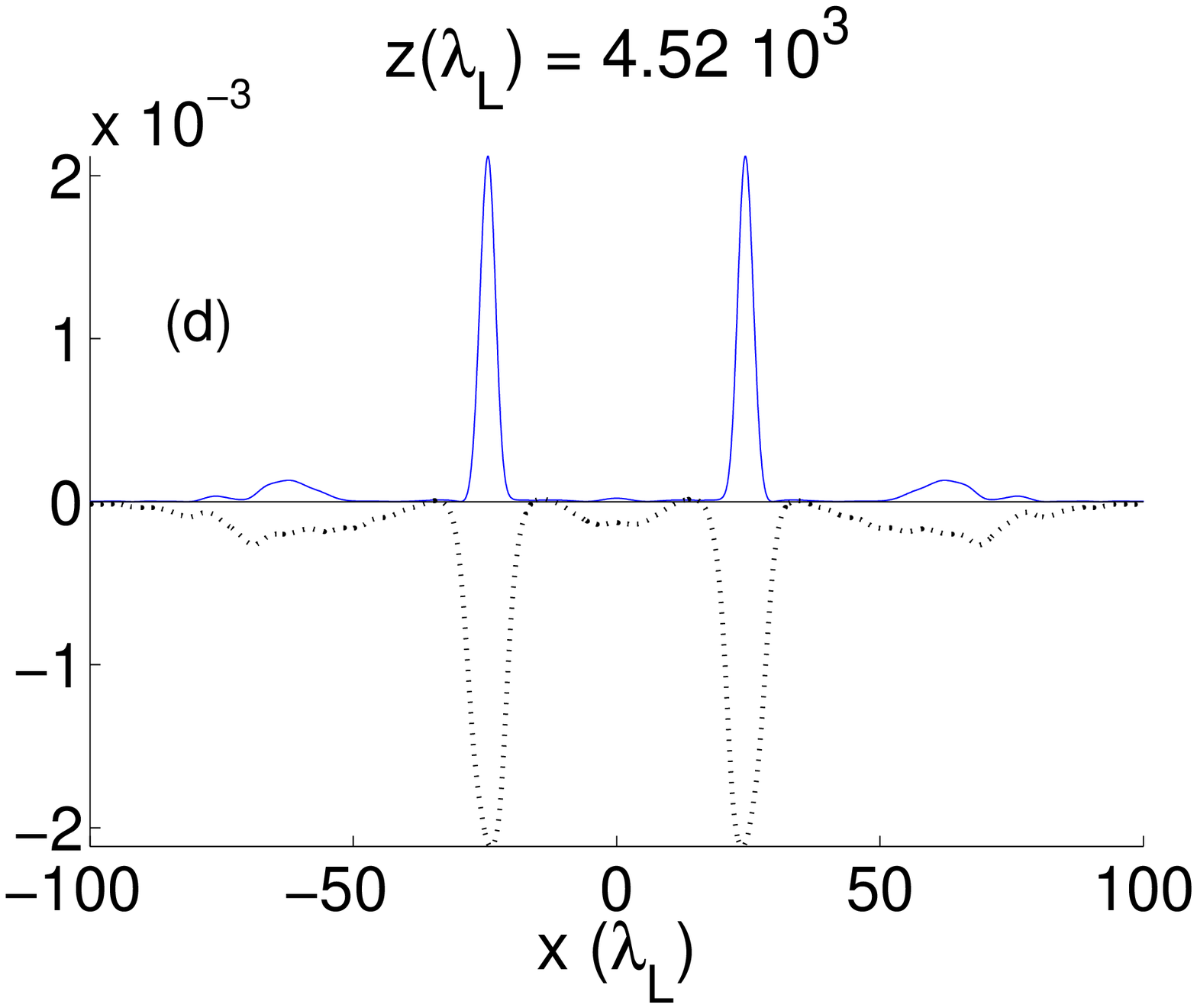}\\
\includegraphics[scale=0.22]{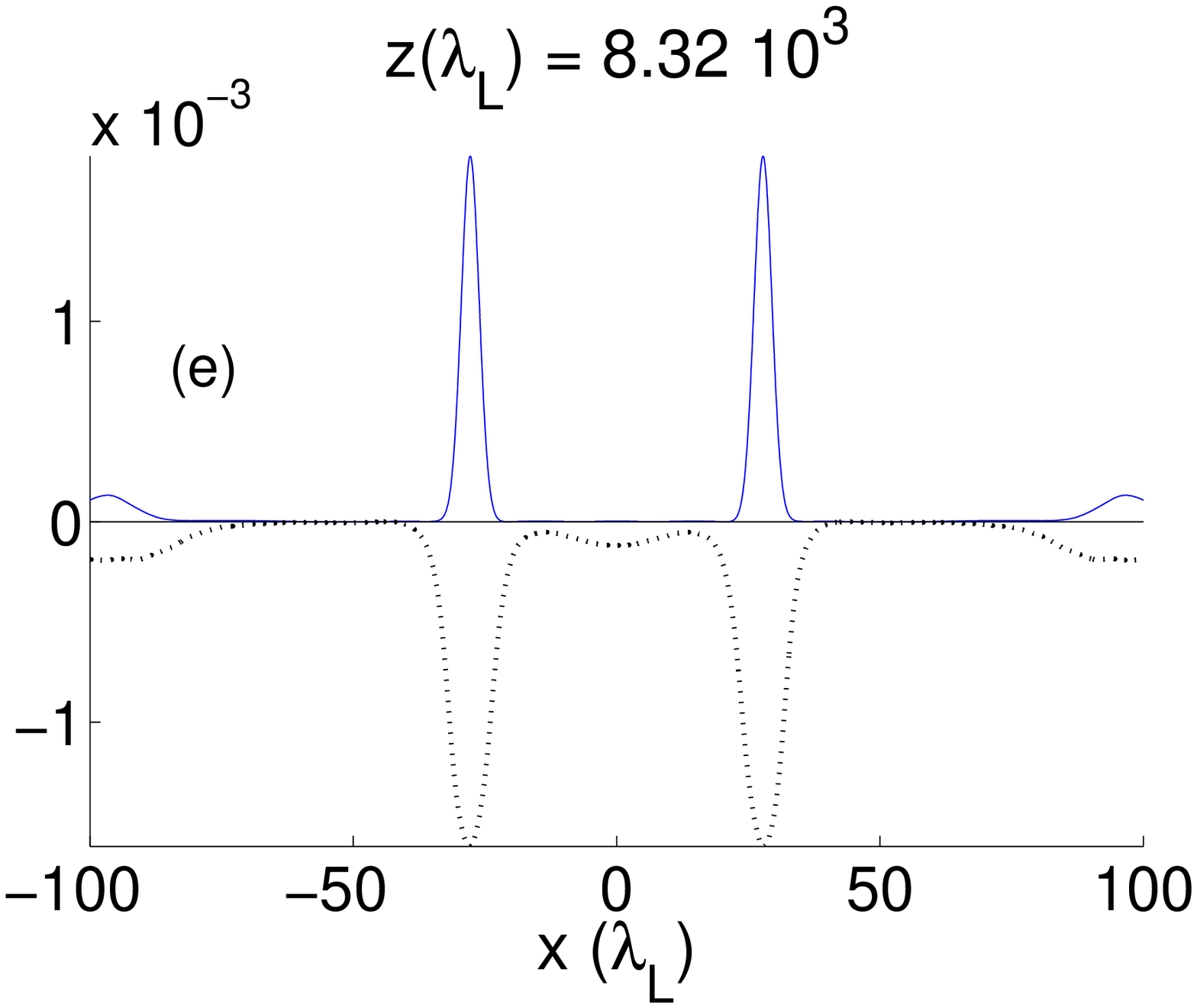}\includegraphics[scale=0.22]{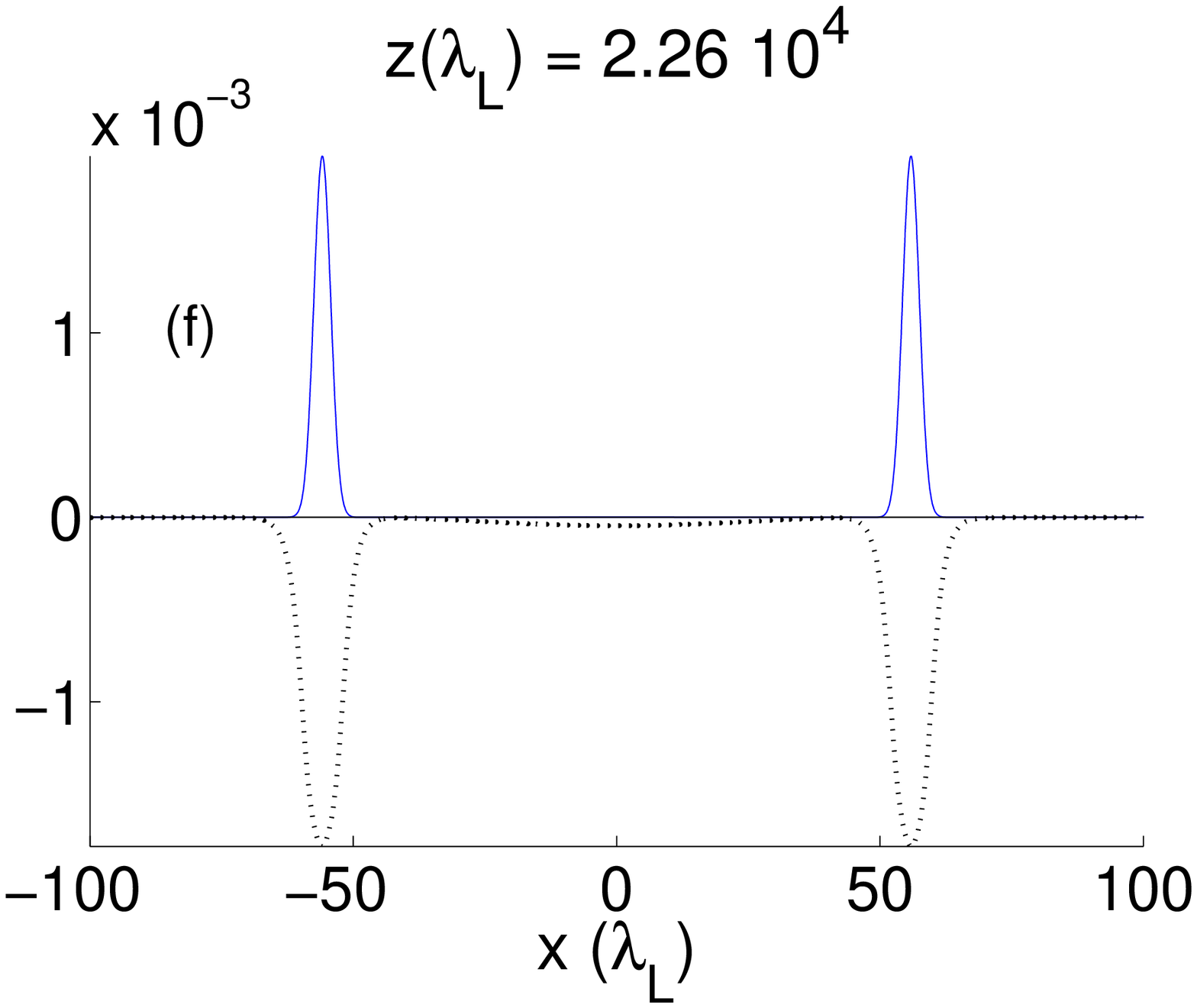}
\caption{Structure emission for  
  $\psi_0 = 0.092$ ($n_0 = 7.62 \,
  10^{19}\textrm{m}^{-3}$).  Solid line: atom wave function, dotted line:
  laser-induced potential acting on the
  atoms (divided by 10 to make the figure more easily readable). Propagation distance as indicated on the
  plots. All quantities normalized as in the text. 
}  \label{fig:run3105900}
\end{figure}

As a consequence of the initial evolution stage, provided the strenght of the focusing
dipole-dipole interaction and that of the defocusing collisional
nonlinearity are initially not completely out of balance, some atoms start to
broaden away from the central structure while a  part of the initial distribution remains
trapped there. The radiation reacts to this process because of the
dependence of the refactive index on the density profile and part of
it  is focused around the peak of the atom density. However, if the
trap induced by the laser is much wider than the atom wave function,
the atoms lost from the central core can still be trapped. What
initially was a hump of disperding 
atoms, can get trapped in a secondary self-generated potential well
and induce mutual localization on the wings. The generation of these
secondary mutually localized structures, keeping the initial laser
with and peak intensity fixed,  should depend on having enough atoms
escaping from the central peak since the escaping atoms must affect
the laser wings to provoke the formation of the secondary
trap. Therefore we have numerically studied the coupled evolution of
(\ref{eq:initialpsi}) and (\ref{eq:initiala}) for
fixed $d_a = 5\lambda_L, d_L=8 d_a$ and fixed $a_0 = 0.1346$ corresponding to 
an initial peak laser intensity of $0.0153 $ mW/cm$^2$,  but varying
$\psi_0$. In the simulations we have $\beta \simeq 38$ corresponding
for instance to a detuning of $100$ times the decay rate for $^{87}$Rb
atoms and $s = +1$.\\
As anticipated, for low $\psi_0$, only a central density peak
remains, a phenomenon studied in \cite{ref:us1}. The central atoms affect the laser profile which creates a
trapping potential. The atoms that escape this trap are not enough
to modify the natural evolution of the laser wings which undergo
well known modulations before diffracting away. Increasing $\psi_0$,
the central structure generated by the system will be obviously
modified, the balance of repulsive collisional interactions and
attractive dipole forces has changed. Furthermore, the effect of
escaping atoms becomes stronger to the point that the same trapping
mechanism can now be realized on the sides of the central
peak. Fig.\ref{fig:run30020019_3102130301}
shows the results of such an interaction for two values
of $\psi_0.$ The laser, as one would expect,  forms analogous localized structures in
correspondence of the atom density peaks.  There is actually a formation of localized
structures even for low $\psi_0$, the very low density of escaping
atoms  can focus extremely weak laser peaks, the process creates
continuous families of mutually localized solutions. However, for low
$\psi_0$ they are hardly visible.
What is interesting about these structures is their fate. They are
 self-consistently formed due to the effect they have on the laser
radiation. Atoms focus the radiation, the radiation in turns exerts a focusing
action on the atoms counterbalanced by their own defocusing
interaction and their kinetic energy. During the initial transient,
which lasts until the atom-laser structures are mutually adjusted to
their own localized form, the lateral peaks are oscillating 
around the point where they have been trapped. They are kept there by the presence
 of the laser trap, laser wings have not yet completely adjusted to
 the newly born structures and they still act as an external trap for
 the atoms. Figs.\ref{fig:run3102130301_details}(a) and (b) show the intermediate
 stage of this transient for the same paramaters as in
 Fig.\ref{fig:run30020019_3102130301}(b). The structures are
 oscillating within the laser-induced trap which is being formed,
 Figs.\ref{fig:run3102130301_details}(c) and (d), 
 and once the laser has completely adjusted  nothing keeps the atom-laser peaks oscillating
 around a fixed position anymore and the structures are free to move
 away, Figs.\ref{fig:run3102130301_details}(e) and (f).  For
 the parameters of Fig.\ref{fig:run3102130301_details}, 
they are ejected from the initial interaction region and
 proceed propagating with constant velocity as solitary-like
 waves. This could be explained by the repulsion due to the central
 peak: the two lateral peaks cannot proceed moving inward because they
 cannot overcome the repulsive barrier due to the central one.(???????)
The position of the lateral peaks as a function of the propagation
distance for the same parameters of
Fig.\ref{fig:run3102130301_details} is shown in
Fig.\ref{fig:run3102130301_jetposition}(a), from which it is clear
that, after an initial transient during which it is quite difficult to
keep track of the structures' positions, 
the two ``jets'' are propagating at constant
velocity. It is also evident how laser and atoms jets move
together. Fig.\ref{fig:run3102130301_jetposition}(b) shows the 
peak value of the atom density of the emitted structures which tend to
stabilize on a stationary value.
In a way, this phenomenon is reminiscent of the emission of solitons
engeneered in nonlinear optics with the aim for instance of
implementing all-optical switching and directional couplers,
\cite{ref:emission}. Whereas in the 
optics case the emission is stimulated only on one side, we obtain two
moving structures because of the symmmetry of the configuration. We
must underline that we refer to the emitted structures 
as solitary-like waves because of their ability to propagate with
unchanged shaped but we have not yet proved their collisional
properties, preliminary results indicate a behaviour strongly
suggestive of a soliton-like nature.\\
The analogy is also suggestive of the possibility of soliton
steering. In fact, the properties of the structures ejected (peak
density, velocity and number of jets) depend on the initial
conditions. Therefore, changing the initial value of $\psi_0$, we have
found jets emitted at different angles with respect to the propagation
direction $z$ and with different peak
densities and peak laser intensities, as can be seen from
Fig.\ref{fig:velocities} which shows jet positions for a few different cases.
This last figure also shows the anomalous behaviour of the structures
emitted starting from $\psi_0 =0.0668$. They initially move clearly
inwards before being ejected. For growing initial peak density, there
seems to be a stronger central trapping capable to attract the lateral
peaks towards the center. Notice from
Fig.\ref{fig:velocities} how, for higher initial $\psi_0$ the jets
tend to be born closer and closer to the central peak, where they are likely
to experience a stronger interaction with it, due to a larger overlap
(compare cases (a) and (b) in that figure with cases (c) and (d) which
have larger $\psi_0$). There is a critical combination of
parameters, which in our case occurs for $\psi_0 = 0.0669$, such that
the two jets are drawn backwards until they collide and fuse at the
center, Fig.\ref{fig:run3102130601}. 
It is known that the result of a collision between two
solitons depending on the relative phase can lead to the fusion of the
two objects, \cite{ref:collision} and references therein, however the nature of the collision
within the model presented here
needs further studies. 
After the merging, the remaining central peak stabilizes and does not
undergo any dynamical changes anymore but it is very likely that such
a structure will not be realized due to the extra-effects that are not
considered within this model and that could play an important role
during the collision.
For higher values of $\psi_0$ no central peak is left while two
lateral peaks are again symmetrically ejected. This could suggest an
instability of the central peak as a
possible explanation of the merging shown by the previous case. If the
central peak is unstable against diffraction/defocusing and the
laser-induced force is not able to keep it trapped, its atoms will
broaden away with two possible outcomes for the jets: Either the
repulsive interaction between the jets and the centrally disperding  atoms is
not strong enough to prevent the jets from merging in the center, or
it is important enough to push them away, compare
Fig.\ref{fig:run3105900} and \ref{fig:run3102130601}.

\begin{figure}
\centering
\includegraphics[scale=0.4]{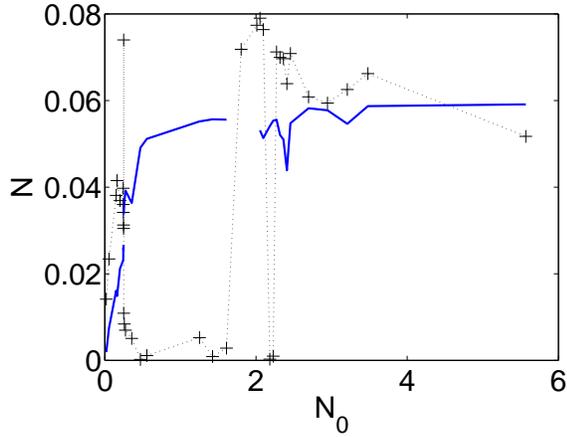}
\caption{Integral of the jets wavefunctions $N =
  \int_{-\infty}^{\infty} |\psi(x, z\rightarrow \infty)|^2 \, dx$
  (solid line) and of the central peak wavefunction (dotted line) versus the integral of
  the initial wavefunction $N = \int_{-\infty}^{\infty} |\psi(x,z=0)|^2 \, dx$. The points were the
solid line is broken correspond to merging and fusion and therefore no
emission of jets at all. All quantities normalized as in the text. 
}  \label{fig:n}
\end{figure}

\begin{figure}
\centering
\includegraphics[scale=0.4]{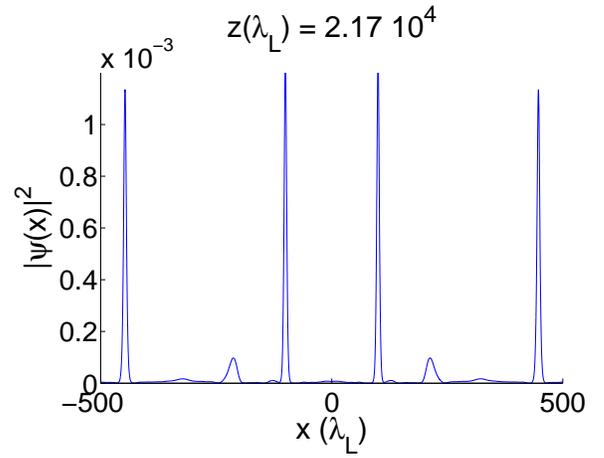}
\caption{Structures emitted for 
  $\psi_0 = 0.2$ ($n_0 = 6.81 \,
  10^{20}\textrm{m}^{-3}$). Propagation distance as indicated on the
  plot. All quantities normalized as in the text. 
}  \label{fig:run3001019}
\end{figure}

It is interesting to notice how the integral of the jets  wavefunction
($N = \int_{-\infty}^{\infty} |\psi(x)|^2 \, dx$ in Fig.\ref{fig:n})
seems to tend to a finite value as a function of the integral of the
initial wavefunction ($N_0$ in Fig.\ref{fig:n}). 
This would be
acceptable from the point of view of soliton behavior: The emitted
solitary-like structures can accomodate a given number of atoms, atoms
in excess will go and form extra jets, an example is shown in
Fig.\ref{fig:run3001019}. 

\begin{figure}
\centering
\includegraphics[scale=0.4]{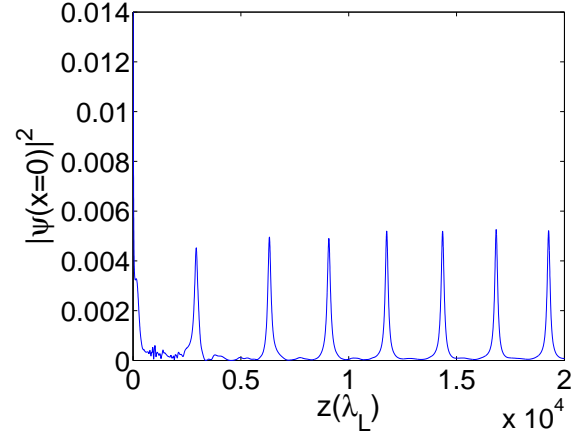}
\caption{Evolution of the central atom density calculated as
  $|\psi(x=0)|^2$ as a function of the propagation distance for
  $\psi_0 = 0.196$ ($n_0 = 6.54 \,10^{20}m^{-3}$).
All quantities normalized as in the text. 
}  \label{fig:run300102049301profile}
\end{figure}

\begin{figure}
\centering
\includegraphics[scale=0.22]{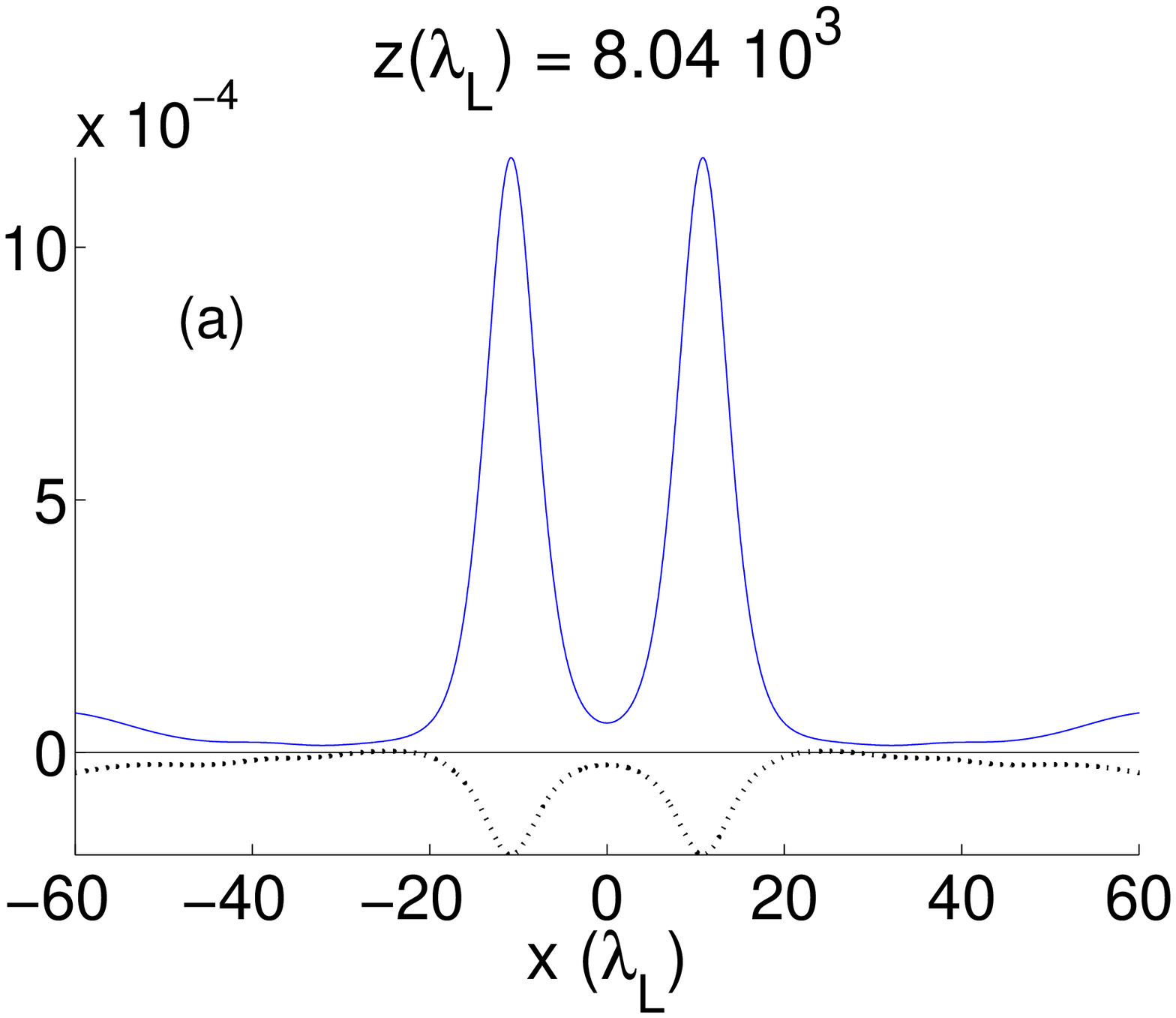}\includegraphics[scale=0.22]{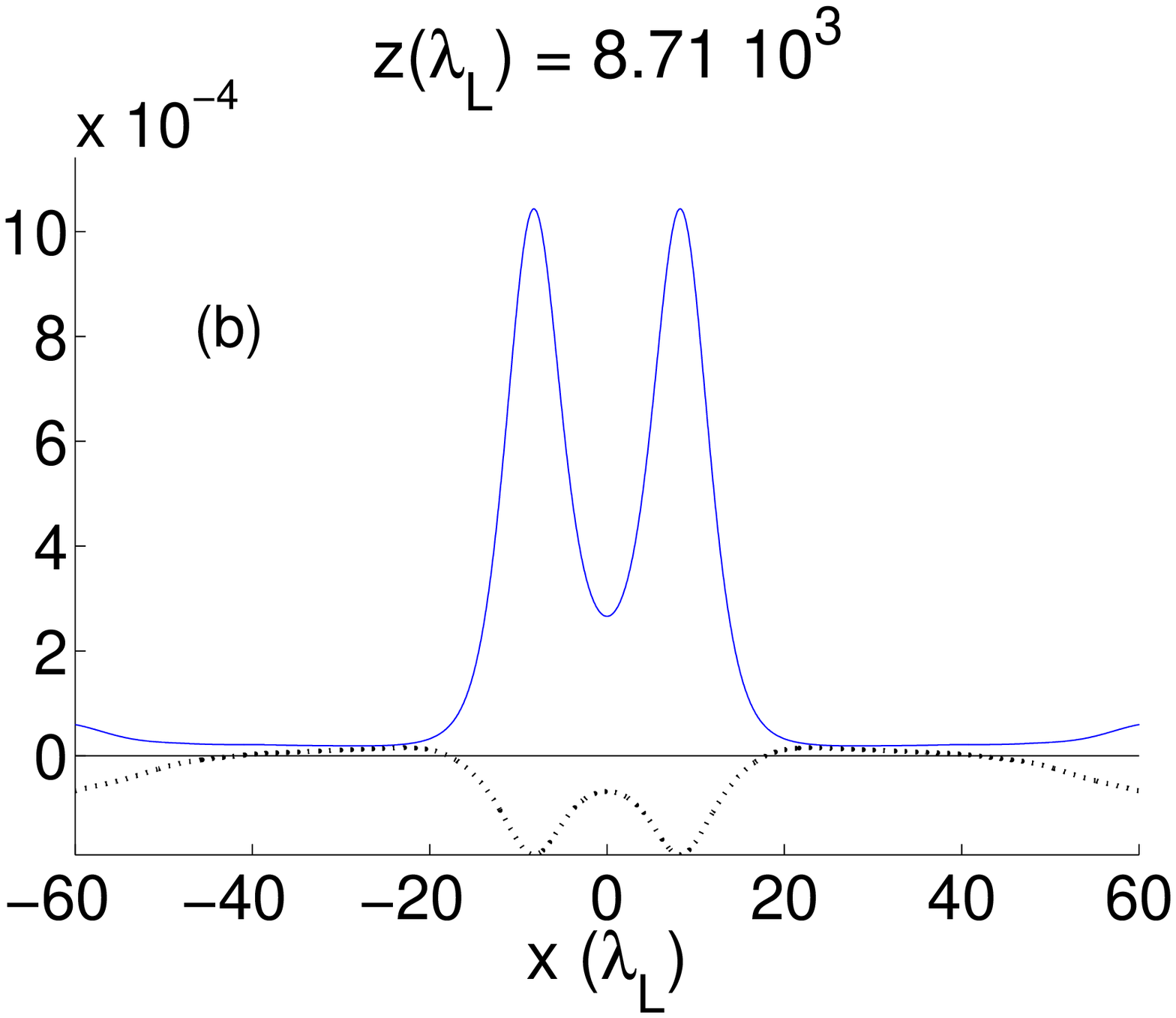}\\
\includegraphics[scale=0.22]{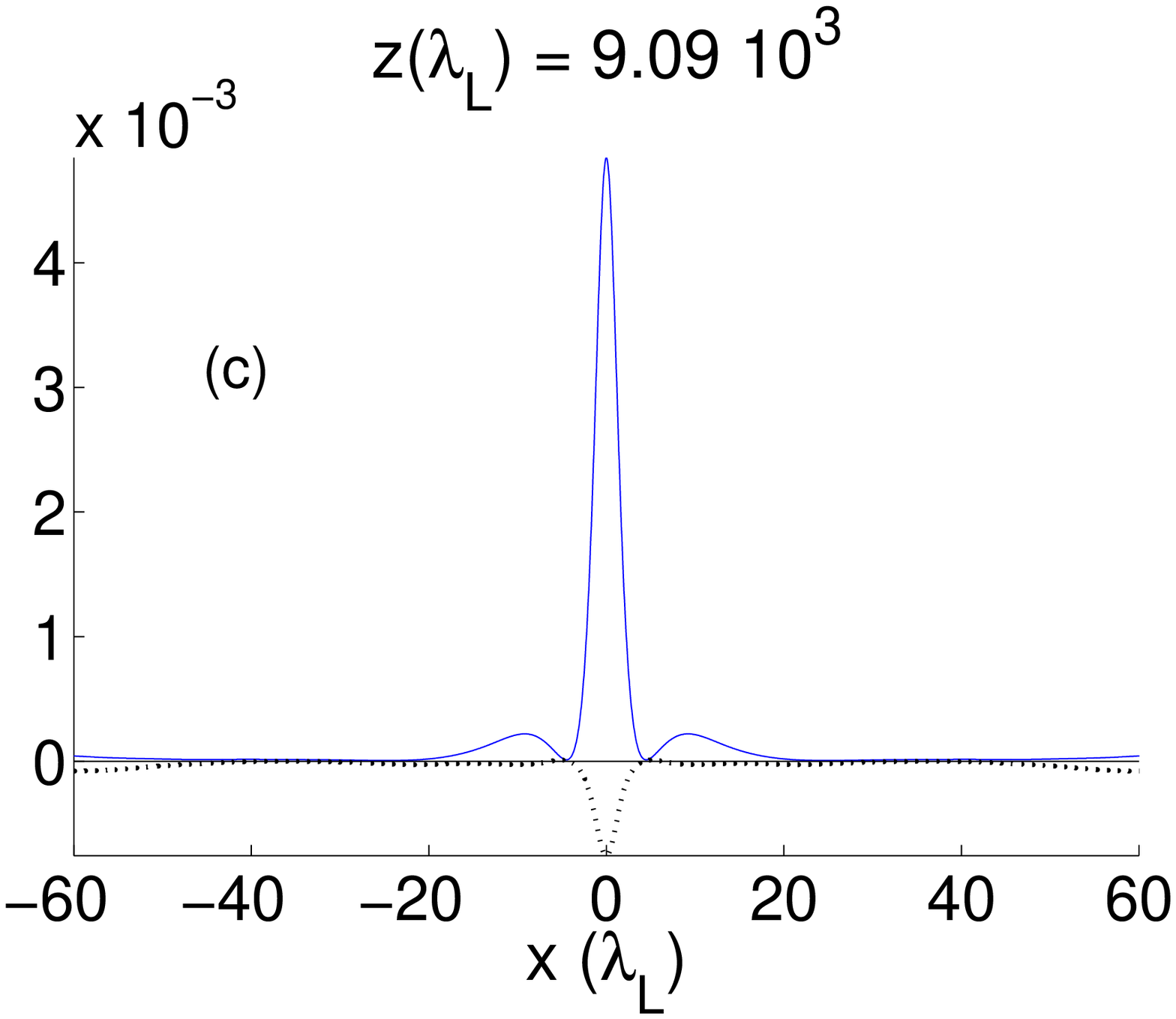}\includegraphics[scale=0.22]{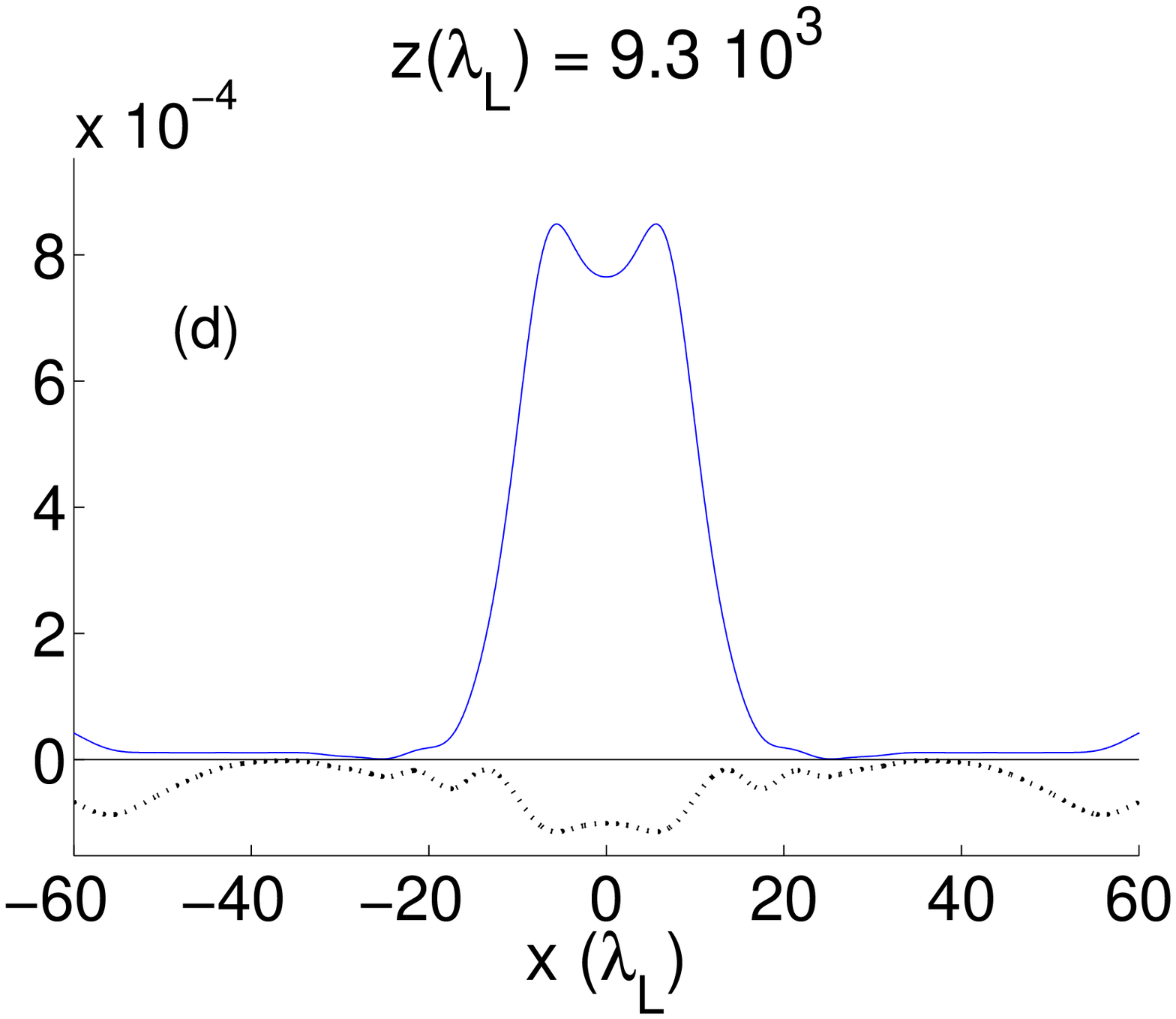}\\
\includegraphics[scale=0.22]{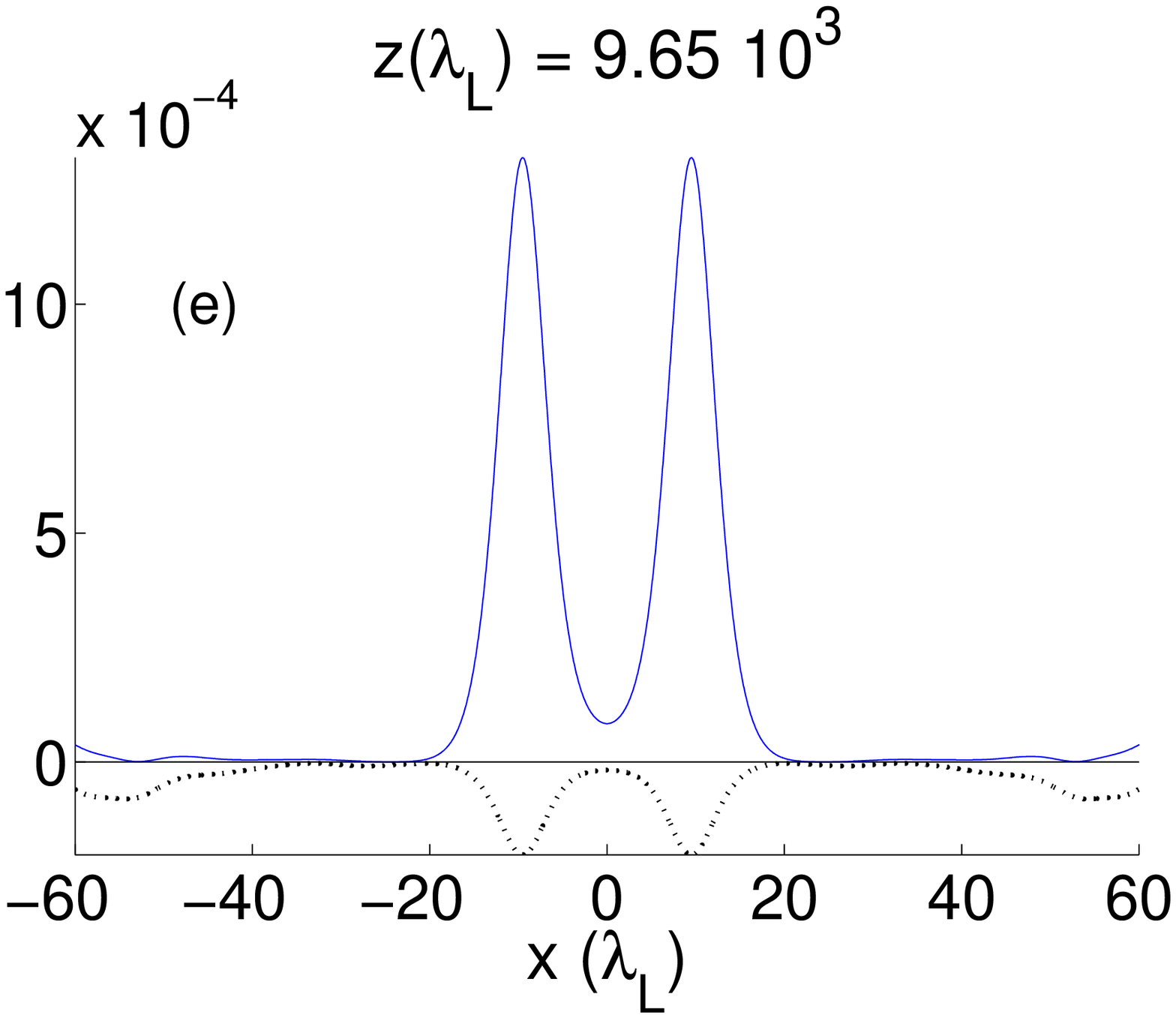}\includegraphics[scale=0.22]{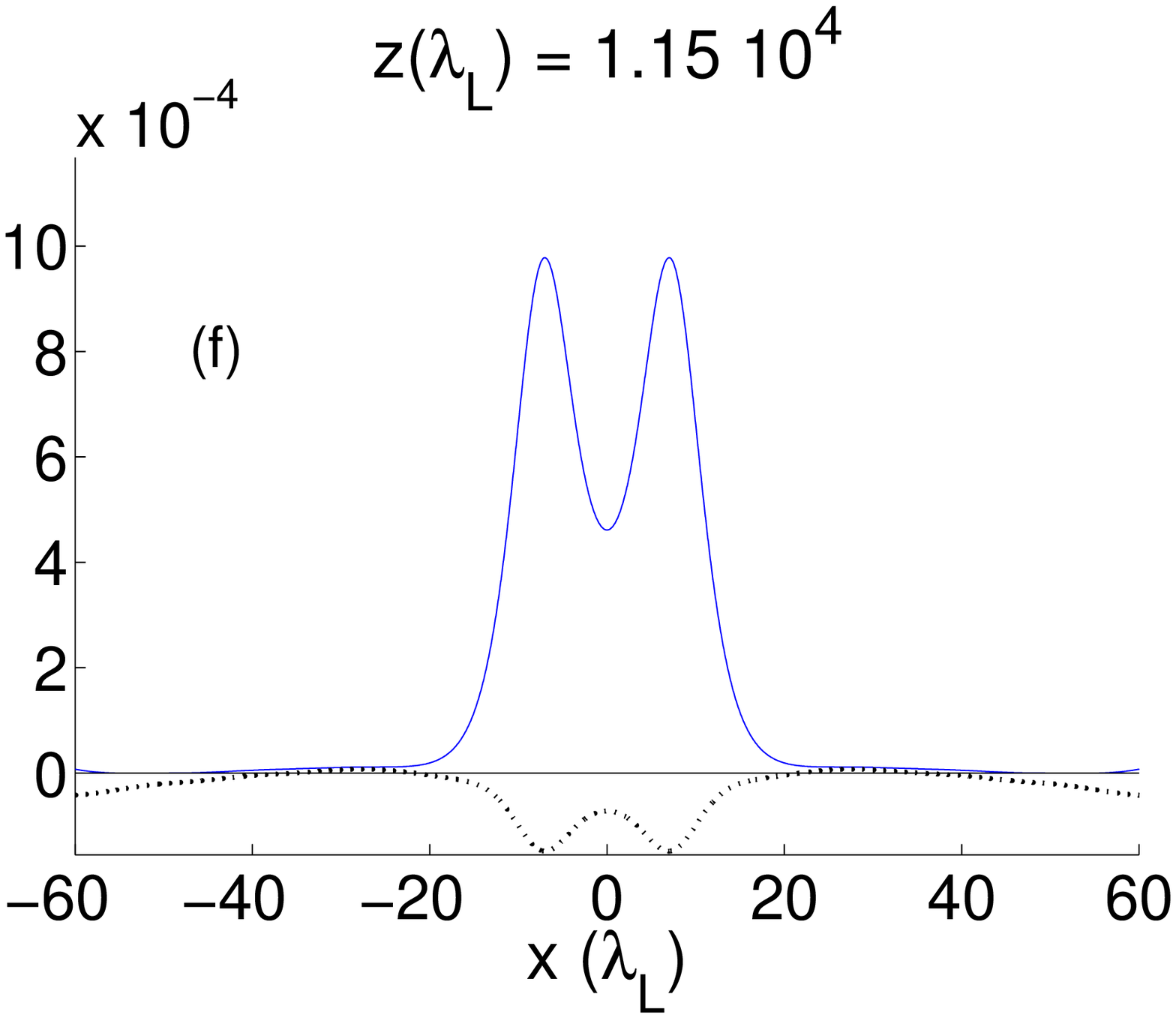}\\
\caption{Snapshots of the evolution of the atom wavefunction for
  $\psi_0 = 0.196$ ($n_0 = 6.54 \,10^{20}m^{-3}$). Propagation
  distance as given in the plots.
All quantities normalized as in the text. 
}  \label{fig:run300102049301}
\end{figure}

A final note concerns one more analogy with an optical soliton
behaviour. It seems in fact possible to excite a structure very
similar to the bound system observed for optical solitons in which two
pulses perform an oscillatory motion by 
bouncing back and forth in their own potential well, \cite{ref:collision}. In a repeated
dance, under particular conditions,
the optical solitons pass through each other, move apart and come to a halt
to move back together. This is what can be seen for a given
choice of initial parameters for the system under
analysis here. Fig.\ref{fig:run300102049301profile} shows 
the value of the atom density at $x=0$ as a function of the
propagation distance for $\psi_0 = 0.196$ and oscillations which would
agree with the presence of a bound soliton 
state are quite evident. 
Corresponding snapshots are  given in Fig.\ref{fig:run300102049301}.\\

\section{Conclusions}

In conclusion, proceeding from the idea that laser-BEC dipole-dipole
interactions can lead to mutually localized structures, we have
analyzed in detail the mechanism of formation of such structures
concentrating on the process through which the structures shed away
the extra atoms and extra radiation.  Numerical simulations seem to
indicate the possibility of generating and emitting secondary
solitary-like wave packets in a jet-like fashon. Although the model
used here is strongly simplified and any comparison with experiment
will require major refinements, the equations we have used enlighten
the main physical effects and it seems possible to choose parameter
regimes in which the effects neglected here will not destroy these
results. This processes could be
a further evidence of the analogy between matter waves and optical
waves and even open the discussion about applications such as soliton stirring in BECs.


\section*{Acknwledgment}
F.C. would like to acknowledge the hospitality of the
department of Radio and Space Physics of Chalmers University of
Technology during the preparation of this work.\\


\end{document}